\PassOptionsToPackage{backref,breaklinks=true,colorlinks,citecolor=blue,urlcolor=[rgb]{0.0,0.0,0.8}}{hyperref}
\documentclass[iop,apj,appendixfloats,twocolappendix]{emulateapj}
\usepackage{adjustbox}
\usepackage{multirow}
\usepackage{colortbl}
\usepackage[fleqn]{amsmath}
\usepackage{amssymb}
\usepackage{upgreek}
\usepackage{enumerate}
\usepackage{float}
\usepackage{placeins}
\usepackage{hyperref}
\usepackage[all]{hypcap}
\accepted{March 6, 2018}

\shorttitle{3D Global Kinetic Pulsar Magnetosphere Models}
\shortauthors{Kalapotharakos et al.} 

\newcommand{\ec}{\epsilon_{\rm cut}}
\newcommand{\ed}{\dot{\mathcal{E}}}
\newcommand{\eacc}{E_{\rm acc}}
\newcommand{\eaccv}{\mathbf{E_{\rm acc}}}
\newcommand{\ir}{\mathcal{F}}

\DeclareMathAlphabet{\mathcalligra}{T1}{calligra}{m}{n}

\begin{document}

\title{3D Kinetic Pulsar Magnetosphere Models:\\ Connecting to Gamma-Ray Observations}

\author{Constantinos Kalapotharakos}
\affil{University of Maryland, College Park (UMCP/CRESST), College
Park, MD 20742, USA} \affil{Astrophysics Science Division,
NASA/Goddard Space Flight Center, Greenbelt, MD 20771, USA}
\email{constantinos.kalapotharakos@nasa.gov}
\email{ckalapotharakos@gmail.com}

\author{Gabriele Brambilla}
\affiliation{Dipartimento di Fisica, Universit\`a degli Studi di
Milano, Via Celoria 16, 20133 Milano, Italy,}
\affiliation{Astrophysics Science Division, NASA/Goddard Space
Flight Center, Greenbelt, MD 20771, USA,} \affiliation{Istituto
Nazionale di Fisica Nucleare, sezione di Milano, Via Celoria 16,
20133 Milano, Italy }

\author{Andrey Timokhin}
\affiliation{University of Maryland, College Park (UMCP/CRESST),
College Park, MD 20742, USA} \affiliation{Astrophysics Science
Division, NASA/Goddard Space Flight Center, Greenbelt, MD 20771,
USA}

\author{Alice K. Harding}
\affiliation{Astrophysics Science Division, NASA/Goddard Space
Flight Center, Greenbelt, MD 20771, USA}

\author{Demosthenes Kazanas}
\affiliation{Astrophysics Science Division, NASA/Goddard Space
Flight Center, Greenbelt, MD 20771, USA}



\begin{abstract}

We present 3D global kinetic pulsar magnetosphere models, where the
charged particle trajectories and the corresponding electromagnetic
fields are treated self-consistently. For our study, we have
developed a cartesian 3D relativistic particle-in-cell code that
incorporates the radiation reaction forces. We describe our code and
discuss the related technical issues, treatments, and assumptions.
Injecting particles up to large distances in the magnetosphere, we
apply arbitrarily low to high particle injection rates and get an
entire spectrum of solutions from close to the
Vacuum-Retarded-Dipole to close to the Force-Free solution,
respectively. For high particle injection rates (close to FF
solutions) significant accelerating electric field components are
confined only near the equatorial current sheet outside the
light-cylinder. A judicious interpretation of our models allows the
calculation of the particle emission and consequently the derivation
of the corresponding realistic high-energy sky-maps and spectra.
Using model parameters that cover the entire range of spin-down
powers of \emph{Fermi} young and millisecond pulsars, we compare the
corresponding model $\gamma$-ray light-curves, cutoff energies, and
total $\gamma$-ray luminosities with those observed by \emph{Fermi}
to discover a dependence of the particle injection-rate, $\ir$, on
the spin-down power, $\ed$, indicating an increase of $\ir$ with
$\dot{\mathcal{E}}$. Our models guided by \emph{Fermi} observations
provide field-structures and particle distributions that are not
only consistent with each other but also able to reproduce a broad
range of the observed $\gamma$-ray phenomenology of both young and
millisecond pulsars.

\end{abstract}

\keywords{pulsars: general---stars: neutron---Gamma rays: stars}



\section{Introduction} \label{sec:intro}
It has been half a century since the first pulsar was observed
\citep{1968Natur.217..709H}. Pulsars are identified as rapidly
rotating Neutron Stars (NS) with huge surface magnetic fields
$B_{\star}$ that reach up to $\approx 10^{14}G$. However, they
continuously lose rotational energy while they radiate over almost
the entire spectrum (form radio to $\gamma$-rays). The corresponding
brightness temperature of radio emission is extremely high
($10^{23}-10^{25}\rm K$) indicating its coherent nature. The exact
mechanism of radio emission remains unknown although it is believed
that it is related to the pair production processes.

However, the radio emission is energetically unimportant in
comparison to the high-energy (e.g. $\gamma$-rays) emission. For
decades we had very limited information about the pulsar
$\gamma$-ray emission and only after the launch of \emph{Fermi} in
2008 we gained gradually access to a plethora of observational data.
Thus, now we have more than 205 detected $\gamma$-ray pulsars (117
of them are compiled in the second pulsar catalog (2PC);
\citealt{2013ApJS..208...17A}) which led to the derivation of
meaningful statistical trends and correlations. This introduced the
\emph{Fermi}-era in pulsar research affecting significantly the
theoretical modeling of the pulsar high-energy emission.

Pulsars are considered to be spherical perfect conductors that
rotate within their own magnetic fields. This rotation polarizes the
charges, supporting an electric field inside the conductor star,
\begin{equation}
\label{eq:cond} \mathbf{E}=-\frac{1}{c}(\mathbf{\Omega}\times
r)\times\mathbf{B}
\end{equation}
where $\Omega$ is the angular frequency of the star. Assuming a
dipole magnetic field and no charges outside the stellar surface the
field structure is provided by the analytic Vacuum Retarded Dipole
(VRD) solution; \citep{deutsch1955}.

Early after the discovery of pulsars it became evident that the huge
voltages due to the accelerating electric field components $\eaccv$,
that exist in VRD solutions, are able to initiate pair cascade
processes \citep{1971ApJ...164..529S}. Actually, it is believed that
the high pair creation efficiency (especially of the
$\mathbf{B}-\gamma$ process) would short-out $\eaccv$ everywhere in
the magnetosphere leading to the so called Force-Free (FF) regime.
Thus, pulsars' radiation can be interpreted as the result of their
(not totally successful) efforts to construct perfect (conductive)
wires using their huge fields and the microphysical processes.

Even though the main principles of the FF solutions had been
described in the late 60's and early 70's \citep[e.g. ][]{GJ69,SW73}
the first realistic FF solution for pulsar magnetospheres was
presented only two decades ago.

\citet*{ckf1999} presented the first FF solution with dipolar
magnetic field for the aligned rotator \cite[$\alpha=0^{\circ}$,
where $\alpha$ is the inclination angle; see also
][]{2005PhRvL..94b1101G,
2006MNRAS.368.1055T,2006MNRAS.367...19K,2006MNRAS.368L..30M,
2012MNRAS.423.1416P,2016MNRAS.455.4267C} and \citet{S2006} for the
oblique rotator \cite[see also ][]{kc2009,2012MNRAS.424..605P,
2013MNRAS.435L...1T}, respectively. These solutions provided not
only the field structure but also the corresponding current and
charge densities and revealed that the current closure is achieved
by the return current, the main part of it flowing along the
equatorial current sheet (ECS) outside the light-cylinder. The ECS
is stable up to large distances \citep*{2012MNRAS.420.2793K} while
its 3D shape is close to the solution for the current sheet of a FF
split monopole solution \citep{bogo99}.

The FF solutions reveal the currents and field structure that make
$\eaccv=0$ everywhere in the magnetosphere. However, due to their
ideal character, they do not provide any information about the
acceleration regions, the corresponding emission, and apparently the
related $\eaccv$. The only way to study the high-energy emission in
FF models was to place arbitrarily the emission regions based on the
pulsar radiation models. \emph{Fermi} observations of the
high-energy cutoff in the spectrum of the Vela pulsar
\citep{2009ApJ...696.1084A}, ruled out polar cap emission models
\citep{1979ApJ...231..854A,1982ApJ...252..337D,1996ApJ...458..278D},
which would produce a super-exponential spectral cutoff from
magnetic pair attenuation. Thus, the candidate accelerating regions
were either (i) along the last open field line, like the Slot Gap
(SG)
\citep{1983ApJ...266..215A,2003ApJ...588..430M,2004ApJ...606.1143M},
or (ii) near the Outer Gap (OG) region
\citep{1986ApJ...300..500C,1996ApJ...470..469R,2001MNRAS.325.1228H},
or (iii) in the ECS
\citep{1996AA...311..172L,2010ApJ...715.1282B,2010MNRAS.404..767C,2012MNRAS.424.2023P,2013AA...550A.101A}
and also (iv) the striped pulsar wind \citep{2005ApJ...627L..37P}.

\citet{2012ApJ...749....2K}, \citet{2012ApJ...746...60L}, and more
recently \cite{2016MNRAS.461.1068C}, motivated by the aforementioned
limitations of the FF solutions, started exploring
resistive/dissipative solutions. These models, using rather simple
macroscopic prescriptions for the current-density $\mathbf{J}$,
regulate the $\eaccv$ through a macroscopic conductivity $\sigma$.
Varying $\sigma$ from 0 to $\infty$ an entire spectrum of solutions
from VRD to FF is produced, respectively. The physical meaning of
$\sigma$ is supposedly related to the local pair multiplicity and
its efficiency to ``kill" $\eaccv$.

These dissipative solutions provided the spatial distribution of
$\eaccv$ and therefore, allowed for the integration of test particle
trajectories in these models and consequently the study of the
corresponding emission. Thus, assuming emission due to curvature
radiation (CR), \citet{2012ApJ...754L...1K} produced the first
$\gamma$-ray light-curves based on dissipative solutions while
\citet{kalap2014} found that a hybrid conductivity model that has
infinite $\sigma$ within the light-cylinder (LC) and high and finite
$\sigma$ outside the LC reproduced the radio-lag $\delta$ vs.
peak-separation $\Delta$ correlation shown in 2PC. In the so-called
FIDO (FF Inside, Dissipative Outside) models, the emission is
produced near the ECS, though the emissivity is not uniformly
distributed.

The successful fit of the 2PC $\delta-\Delta$ correlation gave
confidence that the FIDO macroscopic models provide viable
$\gamma$-ray emission geometries. However, a successful high-energy
emission model should be able to reproduce the observed spectral
properties. Thus, in \citet{2015ApJ...804...84B}, we used the FIDO
models with approximate $\eacc$-values at low $\sigma$ in order to
fit eight bright-pulsars that have published phase-resolved spectra.
This study indicated an increase of $\sigma$ with the spin-down
power $\ed$ and a decrease with the pulsar age.

In \cite{2017ApJ...842...80K}, guided by \emph{Fermi} observations,
we expanded our studies and we refined the FIDO models by
restricting the dissipative regions outside the LC only near the
ECS. Running series of models that covered the entire range of the
observed $\ed$ of Young Pulsars (YP) and Millisecond Pulsars (MP),
and matching the corresponding cutoff energies $\ec$ we revealed a
dependence of $\sigma$ with $\ed$. We found that $\sigma$
increases with $\ed$ for high $\ed$ while it saturates for low
$\ed$. We also found clear indications that the size of the
dissipative zones increase towards low $\ed$ and that the
multiplicity of the emitting particles increase with $\ed$ and
$\alpha$.

\begin{figure*}[!tbh]
\vspace{0.0in}
  \begin{center}
    \includegraphics[width=0.7\linewidth]{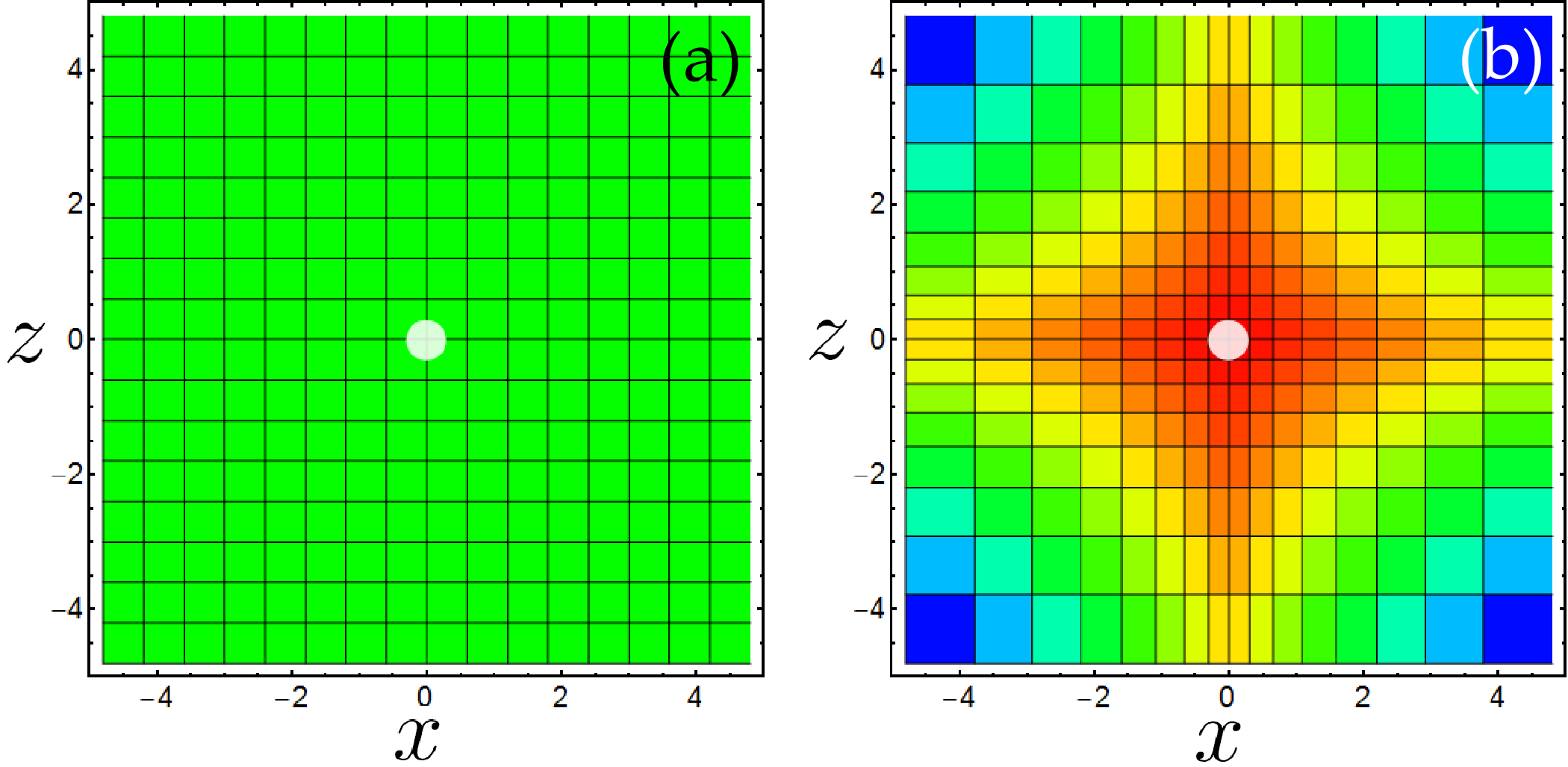}
  \end{center}
  \vspace{0.0in}
  \caption{\textbf{(a), (b)} A 2D slice of the uniform and non-uniform distribution of the
  computational domain implemented in the C-3PA code. The structure shown
  in (b) takes care the load balance issue by taking into account the fact that
  the central regions have much higher particle number densities. Moreover,
  the implementation (b) keeps the original simple cartesian communication between the
  CPUs that ``control" the various sub-domains (i.e. cuboids).
  Even though the implementation (b) is still not optimum it can reduce
  the total computational times, relatively to the implementation (a), by $\sim 1$ order of magnitude.}
  \label{fig:001}
  \vspace{0.0in}
\end{figure*}

Recently, a new approach has been explored to model pulsar
magnetospheres. A few groups attempted to build kinetic models of
pulsar magnetospheres using the Particle-In-Cell (PIC) technique,
when charged particle trajectories and electromagnetic fields are
treated self-consistently
\citep{2014ApJ...785L..33P,2014ApJ...795L..22C,2015MNRAS.448..606C,
2015ApJ...801L..19PA,2015ApJ...815L..19PB,2015NewA...36...37B,
2015MNRAS.449.2759B,2016MNRAS.457.2401C,2017arXiv171003536B}. Some
of these studies
\citep{2014ApJ...795L..22C,2015ApJ...801L..19PA,2015MNRAS.449.2759B}
tried to simulate the pair-creation physics but failed to produce
magnetospheres of low inclination angles filled with plasma close to
the FF ones. One of the problems in these studies is the poor
numerical resolution inherent to any global model which might
prevent an accurate modeling of microphysics of particle
acceleration and pair production. \citet{2016MNRAS.457.2401C}
studied the high-energy emission from their PIC simulations and
found that the main source is synchrotron radiation but with fixed
(with $\ed$) $\gamma$-ray efficiency (not totally consistent with
the observations; 2PC). More recently, \citet{2017arXiv170704323P}
agreed that higher particle multiplicities (i.e. higher $\sigma$)
from pair creation in the outer magnetosphere is needed at higher
$\ed$ values to explain the dependence of the total $\gamma$-ray
luminosity on $\ed$ (i.e. $L_{\gamma}\propto \ed^{1/2}$).

In the present paper, we present our PIC code and we make the first
step by applying ad-hoc (not physically motivated) particle
injections which provide at least field configurations and particle
distributions that are consistent with each other. The main goal of
this study is to explore the role of the global particle injection
rate $\ir$ in both the development of the field structures toward
the FF regime and the characteristics of the corresponding
high-energy emission. More specifically, taking into account the
constraints imposed by \emph{Fermi} data we reveal a relation
between $\ir$ and $\ed$ that is able to reproduce a broad spectrum
of the observed phenomenology (i.e. $\gamma$-ray light-curves and
spectra).

The structure of the paper is as follows. In
Section~\ref{sec:piccode}, we introduce our PIC code, discuss
technical details, and present some basic tests. In
Section~\ref{sec:simset}, we discuss how we build our models and the
adopted model parameters. In Section~\ref{sec:towtheffsol}, we show
a development of solutions from the VRD to the FF one as a function
of the particle injection rate, $\ir$. In Section~\ref{sec:rescale},
we discuss the model interpretation and the method we use in order
to scale the particle energies as well as the corresponding emitted
photon energies to those of real pulsars. In
Section~\ref{sec:fitfermidata}, we present our main results for two
series of models that cover the observed $\ed$-values of YPs and
MPs. For these models, we show the dependence of the spectral cutoff
energies with $\ir$ which, by taking into account the variation of
the \emph{Fermi} $\ec$, leads to a dependence of $\ir$ on $\ed$.
Finally, in Section~\ref{sec:concl}, we summarize our conclusions
and discuss the future prospects.

\section{A particle-in-cell code: C-3PA} \label{sec:piccode}
We have developed a \textbf{C}artesian \textbf{3}D \textbf{P}IC code
for \textbf{A}strophysical studies (hereafter C-3PA). Our code
follows the well known, simple but also powerful algorithm:
\begin{enumerate}
    \item Integrate the time dependent Maxwell's equations one time-step
    using the current provided by the particle collective motions.
    \item Integrate the equations of motion one time-step for all the particles
    taking into account the field structure.
    \item Goto 1.
\end{enumerate}

\begin{figure*}[!tbh]
\vspace{0.0in}
  \begin{center}
    \includegraphics[width=1.0\linewidth]{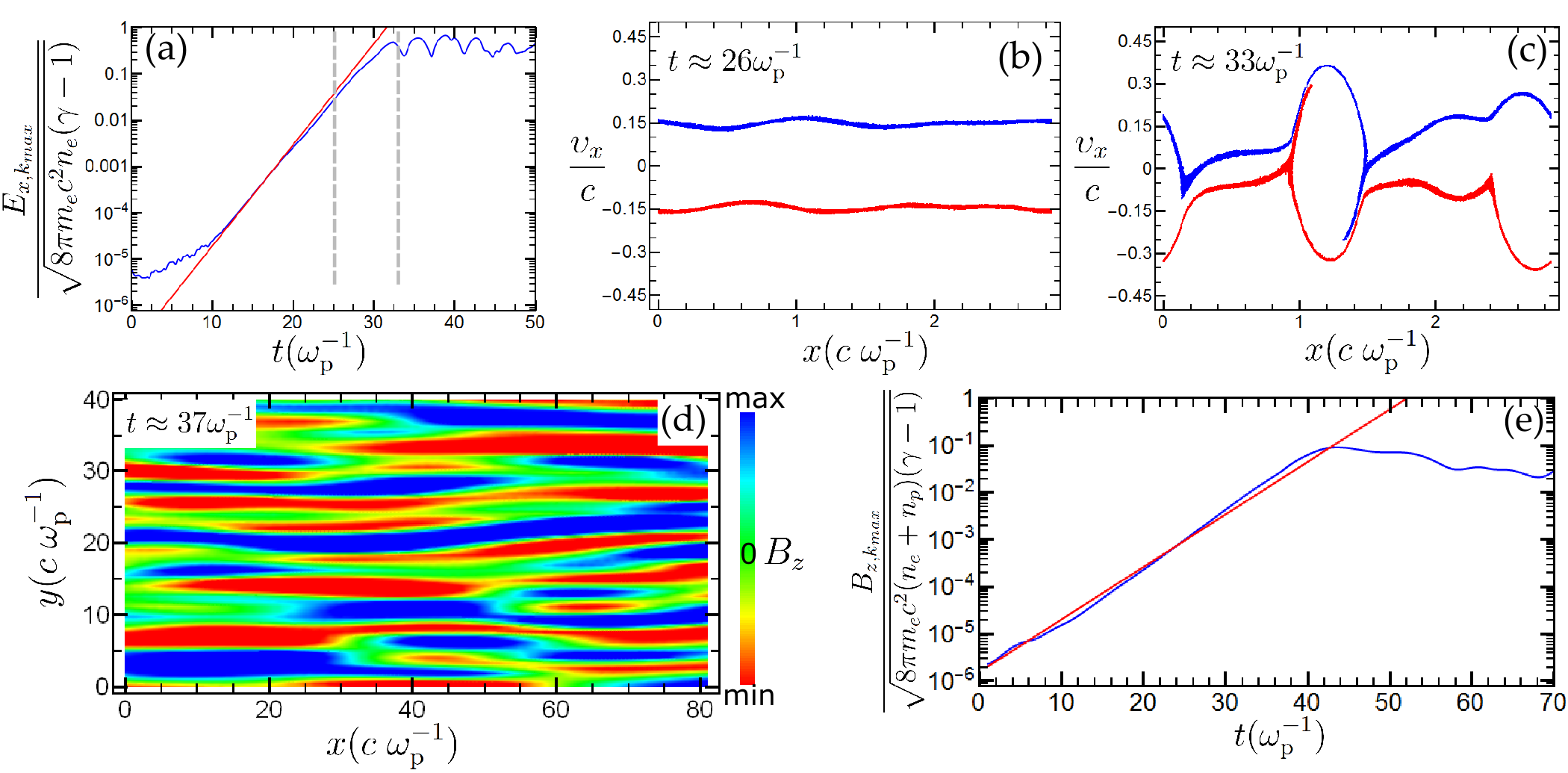}
  \end{center}
  \vspace{0.0in}
  \caption{\textbf{Top row:} The two stream instability test for C-3PA code.
  In \textbf{(a)} the blue line shows the simulated maximum electric
  field component along the charge motion direction as a function
  of time. The red solid line indicates the theoretical growth rate.
  We see that the simulation behaves as theoretically expected.
  In \textbf{(b)} and \textbf{(c)} we plot the phase space
  portraits corresponding to the snapshots indicated
  by the vertical dashed gray lines in \textbf{(a)}.
  \textbf{Bottom row:} {The Weibel instability test for C-3PA code.
  In \textbf{(d)}, the filamentary structure of the magnetic field component $B_{z}$
  at $z=0$ and for the indicated snapshot. \textbf{(e)} The blue line shows
  the time-evolution of the fastest growing mode (i.e. $k_{\rm max}$) of the
  $B_{z}$ component that corresponds to the simulated data. The red line denotes
  the theoretical growth rate of the $k_{\rm max}$ mode \citep{1999ApJ...526..697M}.}}
  \label{fig:002}
  \vspace{0.0in}
\end{figure*}

Our code is written in Fortran 90 and is fully parallelized through
Message Passing Interface (MPI). The code is charge conservative and
in the current version it uses the so called cloud-in-cell (CIC)
scheme for the shape function (i.e. the shape of each macroparticle
is a cube of equal size with one grid cell of the lattice; see
\citealt{1992CoPhC..69..306V}). For the particle mover we
implemented Vay's algorithm \citep{2008PhPl...15e6701V}, which, in
general, provides better accuracy for the drift motion of
relativistic particle trajectories (compared to the standard Boris
algorithm; see \citealt{1991ppcs.book.....B} and references
therein). The forces exerted on the particles are weighted over the
individual cubic particle shapes. The field solver integrates the
time dependent Maxwell's equations through a
Finite-Difference-Time-Domain (FDTD) method applied in a Yee
staggered mesh. In the outer boundary we use a Perfectly Matched
Leyer \citep[PML;][]{1996JCoPh.127..363B,kc2009} that absorbs the
outgoing electromagnetic waves and minimizes the inward reflections.
PML is applied outside a central cubic domain with side $\approx
8.6R_{\rm LC}$ (where $R_{\rm LC}$ is the light-cylinder radius) and
its width is $\approx 0.5R_{\rm LC}$. The particles are removed from
the simulation after they enter the outer layer. They are also
removed when their entire shape enters the central spherical
conductor (rotating NS). Moreover, a gaussian kernel can be applied
for the current density that mimics a higher order shape function
reducing considerably the noise level.

Inside the stellar surface the electric field is always defined by
Eq.~\eqref{eq:cond}. The problem of this configuration is that it
does not incorporate the current closure of the charge carriers that
reach the stellar surface. This leads to the charging of the surface
mostly in the regions the polar-cap passes through \citep[see
also][]{2002ASPC..271...81S}. The problem is more prominent for low
$\alpha$-values because the corresponding polar-caps do not
significantly change their position on the spherical surface.
\citet{2014ApJ...785L..33P} in order to resolve this problem,
simulated the behavior of the spherical conductor by enforcing
charge motions inside the conductor that restore/sustain the
electric field of Eq.~\eqref{eq:cond}. {We have implemented similar
methods where the particles in a thin layer at the stellar surface
are enforced to move in a way that restores the conductive electric
field and the continuity of the parallel (to the stellar surface)
electric field component. Nonetheless, in our simulations, we have
adopted a much simpler and less computationally expensive method
that provides very similar results.} Namely, we integrate Maxwell's
equations down to the stellar radius while at the end of each
time-step we enforce $\mathbf{E}$, within a thin layer ($\sim 3$
cells wide) outside the stellar surface, to go to the conductive
value (i.e. Eq.~\ref{eq:cond}). This treatment ``cleans'' the traces
of the charges that enter the stellar surface and it actually mimics
the current that maintains the electric field in the conductor. In
this approach, the outer surface of this thin layer determines the
effective stellar surface.

We have also incorporated into the particle equations of motion the
radiation reaction forces. The complete expression reads
\citep{1971ctf..book.....L}
\begin{equation}
\label{eq:frr}
\begin{split}
    \mathbf{F_{rr}}&=\frac{2q_e^4}{3m_e^2 c^4}\!\left[\mathbf{E}\!\times\!
    \mathbf{B}\!+\!\mathbf{B}\!\times\!\left(
    \mathbf{B}\!\times\!
    \mathbf{\frac{\mathbf{v}}{c}}
    \right)
    \!+\!\mathbf{E}\left(
    \frac{\mathbf{v}}{c}\!\cdot\!\mathbf{E}
    \right)
    \right]\\
    &-\frac{2q_e^4 \gamma_{\rm L}^2}{3m_e^2 c^5}\mathbf{v}\!\left[\left(\mathbf{E}\!+\!\frac{\mathbf{v}}{c}\!\times\!
    \mathbf{B}\right)^2\!-\!\left(\mathbf{E}\!\cdot\!\frac{\mathbf{v}}{c}\right)^2\right]\\
    &+\frac{2q_e^3 \gamma_{\rm L}}{3m_e c^3}\!\left[\!\left(\!\frac{\partial}{\partial
    t}\!+\!\mathbf{v}\!\cdot\! \mathbf{\nabla}\!\right)\!\mathbf{E}\!+\!\frac{\mathbf{v}}{c}\!\times\!\left(\!\frac{\partial}{\partial
    t}\!+\!\mathbf{v}\!\cdot\! \mathbf{\nabla}\!\right)\mathbf{B}\!\right]
\end{split}
\end{equation}
\vspace{0.1in}

\noindent where $q_{\rm e}$, $m_{\rm e}$ are the electron charge and
mass, respectively and $\mathbf{v}$, $\gamma_{\rm L}$ are the
corresponding particle velocity and Lorentz factor. The third term
involves the convective derivatives of the fields and besides the
implementation difficulty especially in an evolving system we have
found (by applying it to stationary magnetosphere solutions) that it
is negligible compared to the first two terms \citep[see
also][]{2016MNRAS.457.2401C,2010NJPh...12l3005T}. Thus, in C-3PA we
have incorporated only the first two terms following the numerical
scheme suggested in \citet{2010NJPh...12l3005T}.

\begin{figure*}[!tbh]
\vspace{0.0in}
  \begin{center}
    \includegraphics[width=0.85\linewidth]{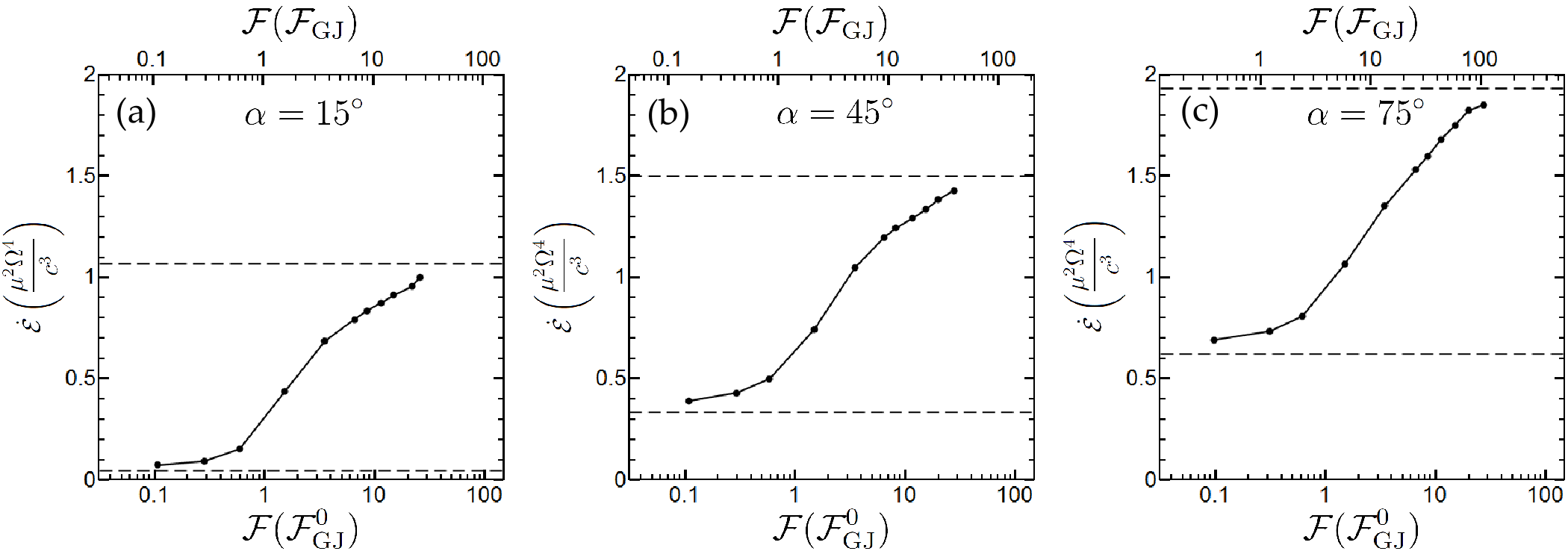}
  \end{center}
  \vspace{0.0in}
  \caption{The spin-down power, for the indicated $\alpha$ values,
  as a function of the global particle injection rate, $\ir$, in log-linear plots.
  On the bottom horizontal axis $\ir$ is presented in units of $\ir_{\rm GJ}^0$
  (i.e. the particle injection rate that corresponds to the GJ flux from the two
  polar caps of the aligned, $\alpha=0^{\circ}$, rotator) while on the top horizontal
  axis $\ir$ is presented in units of $\ir_{\rm GJ}$
  (i.e. the particle injection rate that corresponds to the GJ flux from the two
  polar caps of the corresponding $\alpha$ value). In each panel the
  two dashed horizontal lines indicate the VRD (lower) and FF (higher) $\ed$ values,
  respectively.}
  \label{fig:003}
  \vspace{0.0in}
\end{figure*}

In almost every PIC code the run time is determined mainly by the
work load that corresponds to the integration of the particle
equations of motion (and not the integration of Maxwell's
equations). In our problem the spatial particle distribution is
quite non-uniform. The particle number density outside the stellar
surface is much higher than the one at the edges of the
computational domain. Assuming an MPI implementation where all the
central processing units (CPUs) ``control'' equal computational
volumes (Fig.~\ref{fig:001}a) the performance is slower (in several
cases by more than one order of magnitude) compared to the
theoretical one that corresponds to the total number of operating
CPUs. This happens because the CPUs that ``control'' the outer parts
of the domain have to wait for the inner CPUs to finish the
calculations for relatively much higher numbers of particles. C-3PA
takes care of this load balance issue by implementing a rather
simple non-uniform volume distribution (Fig.~\ref{fig:001}b) that,
in general, increases the controlled volume with the distance from
the center. The advantage of the scheme shown in Fig.~\ref{fig:001}b
is that it keeps a straight-forward communication between the
neighbor cuboids. A proper choice of the different length sizes of
the rectangular cuboids (rectangles in Fig.~\ref{fig:001}b) improves
considerably the computational speed. In particular, for simulations
with $16^3=4096$CPUs, we gain a factor $\sim 7$.

Besides all trivial tests, we have reproduced the two stream
instability \citep{1991ppcs.book.....B}. It is well known that two
sets of opposite streams of charged particles are unstable. Any
small charge imbalance (i.e. perturbation) creates an electric field
along the motion direction that grows exponentially. {In
Fig.~\ref{fig:002}a we show, for a configuration of initially two
opposite streams of $e^{-}$ within a proton background, the
evolution of the amplitude of the fastest growing mode of the
electric field component $E_{x}$ along the motion direction $x$.}
The blue line corresponds to the simulated data while the red line
indicates the theoretical growth rate which is equal to $\omega_{\rm
p}/2$ where $\omega_{\rm p}$ is the electron plasma frequency.
Figure~\ref{fig:002}b,c show the phase space for the indicated times
(see also the dashed light gray vertical lines in
Fig.~\ref{fig:002}a) that correspond to a snapshot during the
exponential growth and to a snapshot when the instability has been
fully developed. {Moreover, we have checked Weibel type of
instabilities. In the bottom row of Fig.~\ref{fig:002}, we show the
results corresponding to a 3D simulation the initial state of which
consists of uniformly distributed $e^--e^+$ pairs that have opposite
velocities along the $x$ axis with $\gamma_{\rm L}=15$. The total
net current is 0 since half of the pairs have $e^-$ ($e^+$) moving
along the positive (negative) direction of the $x$ axis while the
other half has $e^-$ ($e^+$) moving along the corresponding negative
(positive) direction. We note that the plasma is cold in the
perpendicular direction ($\gamma_{\rm L_\perp}$=1). In
Fig.~\ref{fig:002}d, we plot, in the indicated color scale, the
values of the component $B_z$ of the magnetic field at
$t\approx22\omega_p^{-1}$, where $\omega_p$ is the fundamental (i.e.
non-relativistic) plasma frequency. In Fig.~\ref{fig:002}e, the blue
line shows the time evolution of the amplitude of the $k_{\rm max}$
mode of the $B_{z}$ magnetic field component where $k_{\rm max}$ is
the wave-number corresponding to the highest growth rate
\citep{1999ApJ...526..697M,2012ApJ...744..182S,2015ApJ...806..165K}.
The red line denotes the theoretical growth rate for the $k_{\rm
max}$ mode. We see that C-3PA reproduces not only the corresponding
filamentary field structure but also the correct growth rate. We
note also that the saturation occurs when $B$ reaches to
subequipartition values $\sim0.1$ \citep{1999ApJ...526..697M}}.

\begin{figure*}[!tbh]
\vspace{0.0in}
  \begin{center}
    \includegraphics[width=0.7\linewidth]{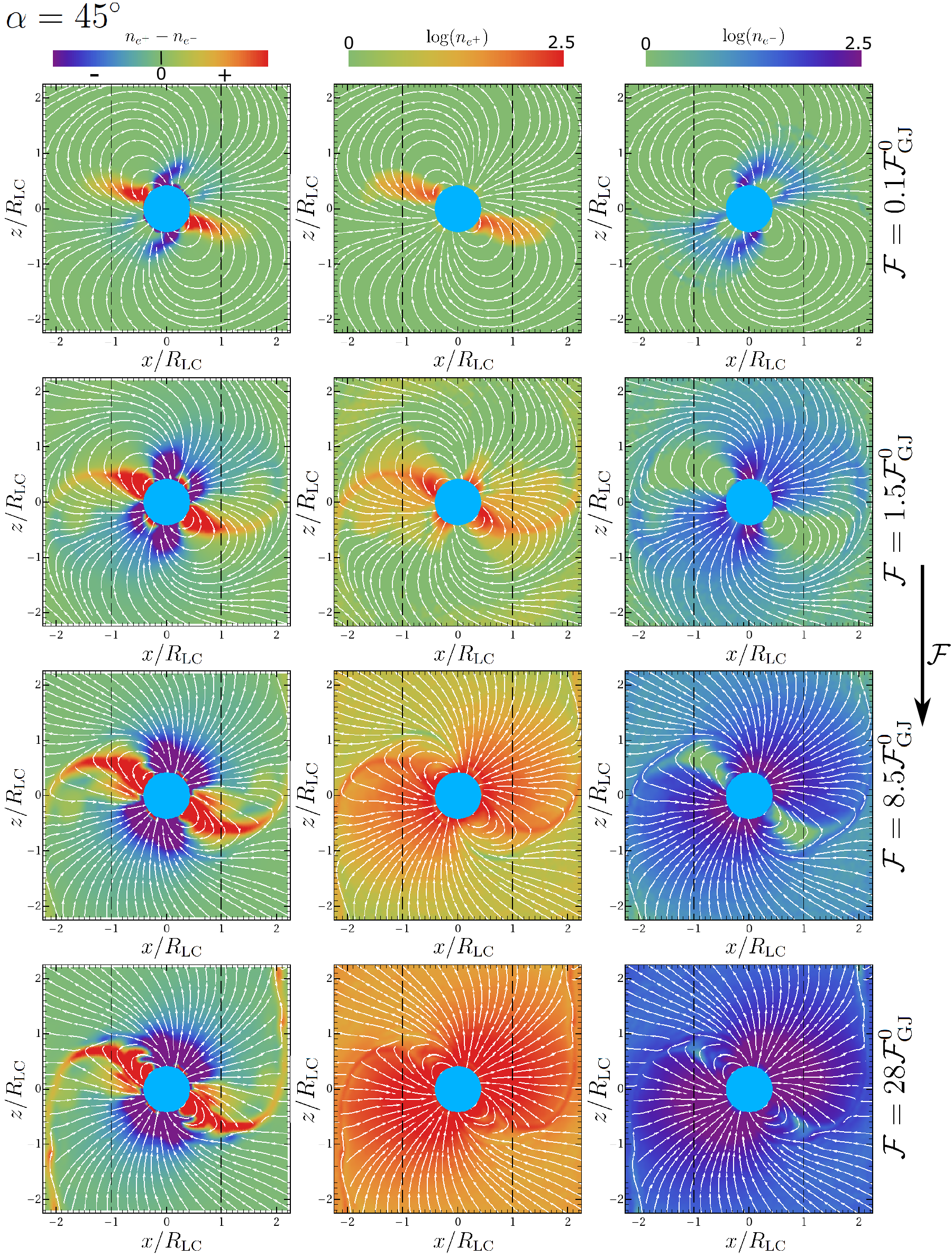}
  \end{center}
  \vspace{0.0in}
  \caption{In the left-hand, middle, and right-hand columns we plot, on the
  $\pmb{\mu}-\pmb{\Omega}$ plane, the charge density, $e^+$ number density,
  and $e^-$ number density, respectively, in the indicated color scales,
  for simulations of  $\alpha=45^{\circ}$. In each panel, the white lines show
  the poloidal magnetic field while each row corresponds to
  simulations for the $\ir$ values indicated in the figure. The vertical black lines
  indicate the LC.}
  \label{fig:004}
  \vspace{0.0in}
\end{figure*}

\section{Simulation setup} \label{sec:simset}

The present study focuses on exploring kinetic (PIC) pulsar
magnetosphere models that neglect the pair creation microphysics. As
mentioned above, the primary goal is to find how the solutions and
the related properties depend on a single parameter, the global
particle injection rate $\ir$. Thus, for the models presented
throughout this paper the particle injection is made based on the
following prescription: at each time-step and at each cell up to
$r=2.5R_{\rm LC}$ one pair ($e^+,~e^-$) is injected at rest
($\gamma_{\rm L}=1$; $\gamma_{\rm L}$ is the Lorentz factor) as long
as the plasma magnetization
\begin{equation}
    \label{eq:magnetiz}
    \upsigma_{\rm M}=\frac{B^2}{8\pi(n_{e^+}+n_{e^-})m_{\rm e}c^2}
\end{equation}
exceeds a locally predefined value, $\Sigma$, that varies according
to
\begin{equation}
    \label{eq:magnthr}
    \Sigma=\left\{
    \begin{array}{ll}
        \Sigma_0 \left(\frac{r_{\rm s}}{r}\right)^3  & \mbox{if } r \leq R_{\rm LC} \\
        \Sigma_0 \left(\frac{r_{\rm s}}{R_{\rm LC}}\right)^3 \frac{R_{\rm LC}}{r} & \mbox{if } r > R_{\rm LC}\\
    \end{array}
\right.
\end{equation}
where $m_{\rm e}$ is the electron mass, $n_{e^-}$ and $n_{e^+}$ are
the $e^-$ and $~e^+$ number densities, $r_{\rm s}$ is the stellar
radius in the simulation, and $\Sigma_0$ is the value of $\Sigma$ at
$r=r_{\rm s}$. The adopted dependence of $\Sigma$ on $r$ is similar
to the $B$ dependence in the FF solutions. Assuming that $n\propto
B$ then $\sigma_{\rm M}\propto B$. Thus, Eq.~\ref{eq:magnthr}
provides ideally the same $\sigma_{\rm M}$ value everywhere. Even
though this is an oversimplification ($n$ is not proportional to $B$
everywhere in the magnetosphere), we use this threshold prescription
because it places appropriate amounts of the injected plasma
(compared to the total/global injection rate) in a large area of the
magnetosphere while avoiding over injection of particles (e.g. in
the closed field line regions). In that sense, Eq.~\ref{eq:magnthr}
is not essential and a somewhat different radial profile would
provide similar results.

\begin{figure*}
\vspace{0.0in}
  \begin{center}
    \includegraphics[width=1.0\textwidth]{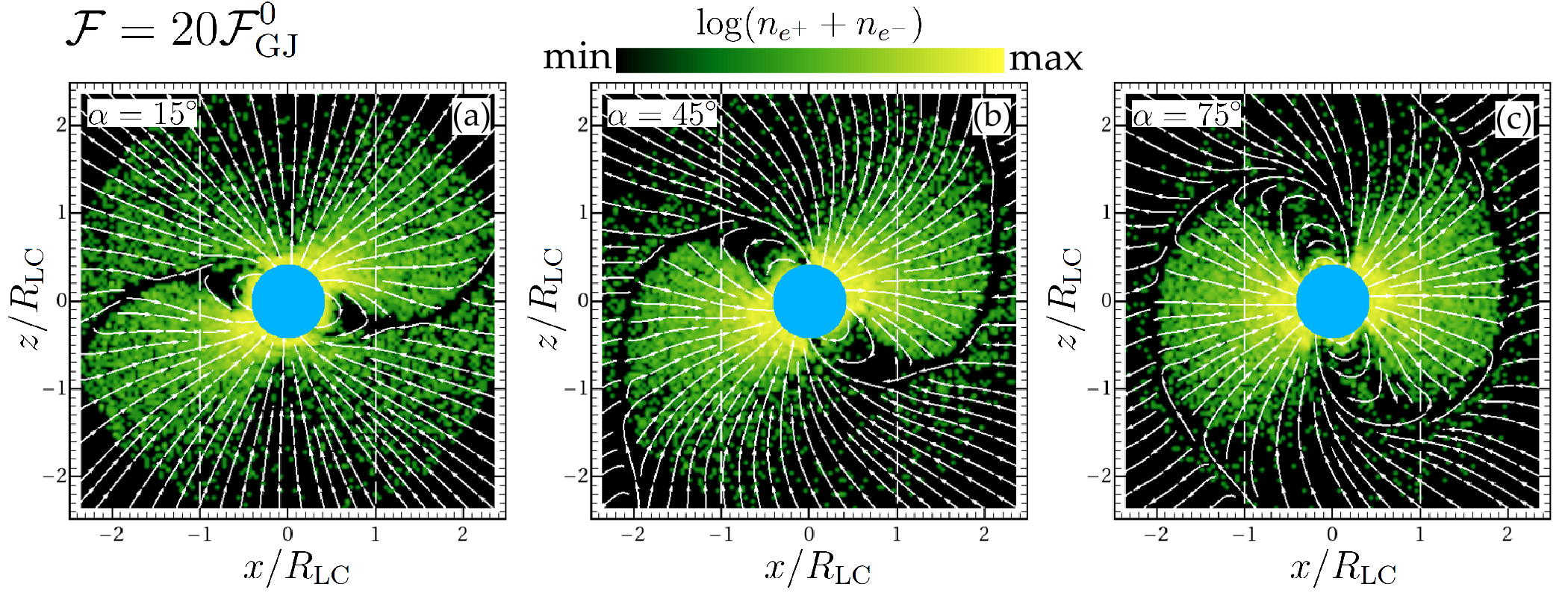}
  \end{center}
  \vspace{0.0in}
  \caption{The number density of the particles that are injected
  per time-unit, in the indicated logarithmic color scale. The
  plotted results are on the $\pmb{\mu}-\pmb{\Omega}$ plane for
  the indicated $\alpha$ values and for $\ir=20\ir_{\rm GJ}^0$. The solid
  white lines indicate the poloidal magnetic field lines while the vertical dashed
  white lines point out the LC.}
  \label{fig:007}
  \vspace{0.0in}
\end{figure*}

Our code starts with some initial (estimated) $\Sigma_0$ value which
is adjusted in time so that $\ir$ reaches the originally adopted
goal-value. The $\ir$ unit corresponds to the Goldreich-Julian flux
from both polar-caps $\ir_{\rm GJ}=\ir_{\rm
GJ}^0\cos(\alpha)$\footnote{This definition breaks down very close
to $\alpha=90^{\circ}$ where a more detailed calculation is needed
that takes into account the particle number density $(n_{\rm GJ})$
distribution on the entire polar cap and not only the value at the
magnetic pole.}, where $\ir_{\rm GJ}^0=B_{\rm s}\Omega A_{\rm
pc}/(\pi q_{\rm e})$ ($q_{\rm e}$ is the elementary electric charge,
$B_{\rm s}$ is the stellar surface magnetic field, and $A_{\rm pc}$
is the corresponding polar cap area) is the $\ir_{\rm GJ}$ value for
$\alpha=0^{\circ}$.

Starting from a configuration close to the VRD, we start injecting
particles based on the aforementioned prescription and let our
simulations evolve for at least 1.5 periods while the steady-state
is achieved after $\sim 1$ period.

The stellar period in the simulation is $P_{\rm s}=0.1\rm s$ which
places the LC at $R_{\rm LC}=4.8\times 10^8\rm cm$. The adopted
magnetic field at $r_{\star}=10\rm km$ (actual NS radius) is $B_{\rm
s}=10^6\rm G$ which allows us to achieve high magnetization values
even for the highest injection rates. For the simulations with the
highest particle injection rates the magnetization is $\sim 500$ at
the stellar surface and $\sim 40$ near the LC and it never falls
below 10 in the regions\footnote{The regions near the ECS where the
local $B$ value decreases dramatically are excluded.} we present our
results. We have also run simulations with $B_{\rm s}=10^5\rm G$ and
the results presented below remained unaffected. If not mentioned
otherwise the simulation resolution presented below is 25 grid
points per $R_{\rm LC}$. The simulating stellar radius is at $r_{\rm
s}=0.28R_{\rm LC}$ (resolved by 7 grid points) while the kernel
layer thickness is 2.5 grid cells which places the effective stellar
radius at $r_{\rm eff}=0.38R_{\rm LC}$. {We note that these stellar
radii are unrealistically high. Nonetheless, many previous studies
\citep{S2006,kc2009,2012MNRAS.420.2793K,
2012ApJ...749....2K,2012ApJ...746...60L,
2013MNRAS.435L...1T,2014ApJ...785L..33P,
2015MNRAS.448..606C,2015ApJ...801L..19PA} have shown that the field
and current structures on the polar caps remain practically
unaffected as long as the stellar radius is not close to the LC.}

The charge $q_{\rm M}$ of the individual macroparticles determines,
for each $\ir$-value, the average number $N_{\rm Mc}$ of
macroparticles per cell. The smaller the $q_{\rm M}$ (i.e. higher
$N_{\rm Mc}$) the lower the noise level (i.e. the field and current
fluctuations). However, we have found that the higher the $\ir$ the
lower the $q_{\rm M}$ should be in order to keep the noise at
relatively low levels. As we also discuss in
Section~\ref{sec:rescale} for solutions near the FF (i.e. high
$\ir$) the actual accelerating electric fields become small and the
noisy fields can become comparable. Apparently, this implies that
for higher $\ir$ we need disproportionately high numbers of
macroparticles which makes the corresponding simulations very
expensive.

The time-step $\Delta t_{\rm s}=4\times 10^{-5}P_{\rm s}$ guarantees
that, for the most dense magnetosphere regions of the models with
the highest injections rates presented in this paper, $(\omega_{\rm
p}\Delta t_{\rm s})^{-1}>6$, where $\omega_{\rm p}$ is the
fundamental plasma frequency $\omega_{\rm p}=\sqrt{4\pi
(n_{e^+}+n_{e^-})q_{\rm e}^2/m_{\rm e}}$. In some cases the adopted
global time-step doesn't resolve the gyro-motion. In these cases,
the particle equations of motion are integrated by smaller
time-steps. Following the orbits finely doesn't affect the currents
that depend only on the motion within the global time-step.
Nonetheless, we have found that it provides more accurate particle
energies and as we discuss in Section~\ref{sec:rescale} allows us to
adjust the particle gyro-motion.

We note that in some cases, the adopted regular spatial resolution
does not resolve the skin-depth $\lambda_{\rm D}=c/\omega_{\rm p}$
for the low $\gamma_{\rm L}$ particles in the most dense
magnetosphere regions. This implies that this particle population
could be artificially heated to energies higher than expected
physically. However, the corresponding thermal $\gamma_{\rm L}$
values are much smaller ($\sim 1$ order of magnitude) than that
corresponding to the full potential drop across the polar-cap (i.e.
$\gamma_{\rm Lmax}=q_{\rm e}B_{\rm s}r_{\star}^3/m_{\rm e}c^2R_{\rm
LC}^2$) while the number of these particles seems to be small and
doesn't affect the particle energy distributions. Actually, we have
found that the particle energy distribution is more sensitive to the
$q_{\rm M}$ and $\Delta t_{\rm s}$ values discussed above.
Nonetheless, we have run simulations for the highest $\ir$-values
doubling the spatial resolution, and the particle energy
distributions and all the results presented in the next sections
remained unaffected.

\begin{figure*}[!tbh]
\vspace{0.0in}
  \begin{center}
    \includegraphics[width=0.76\textwidth]{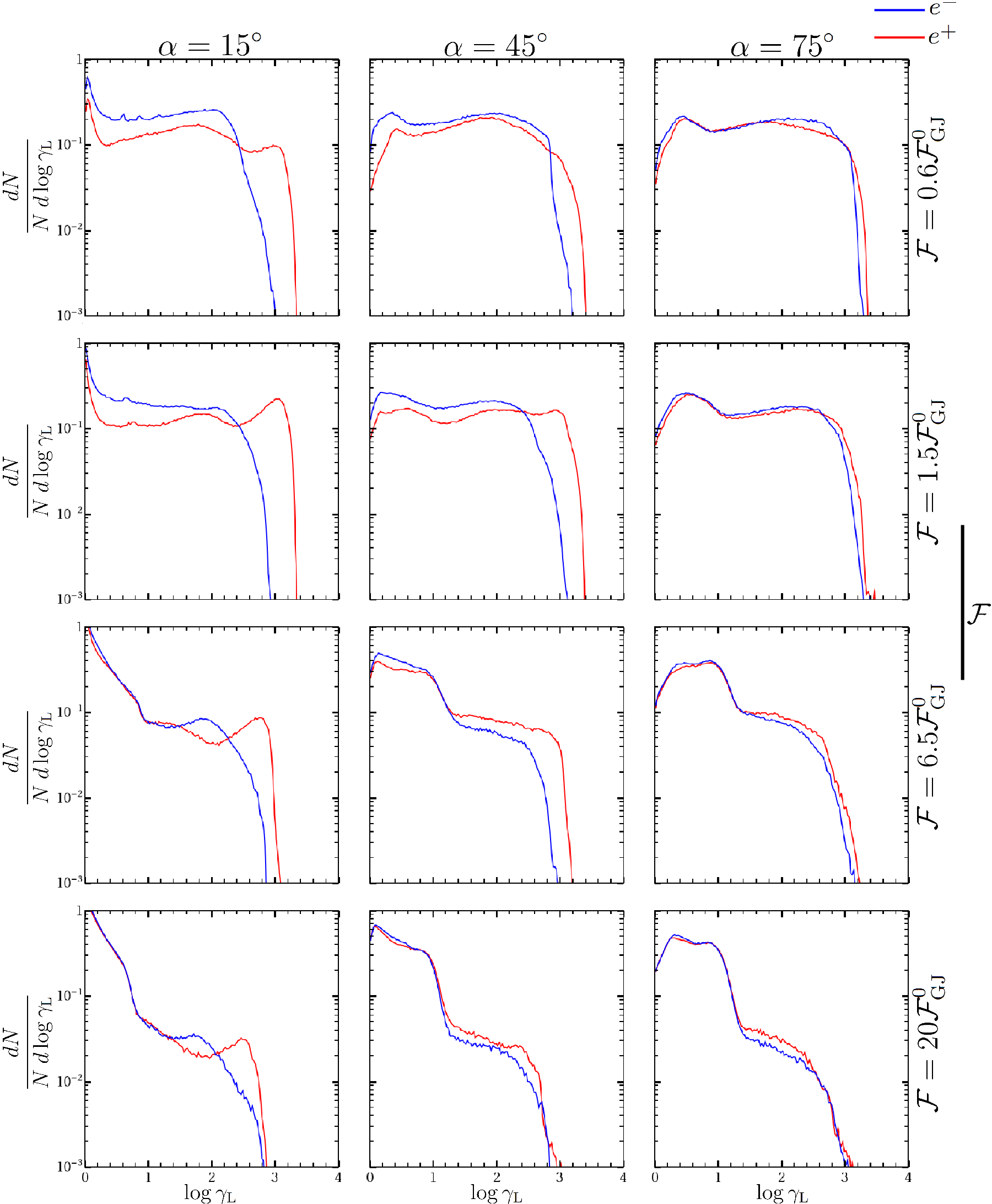}
  \end{center}
  \vspace{0.0in}
  \caption{The normalized $e^-$ (blue)
  and $e^+$ (red) energy distributions. Each column corresponds to the
  indicated $\alpha$ values while each row corresponds to the indicated $\ir$
  values.}
  \label{fig:0045}
  \vspace{0.0in}
\end{figure*}

\section{Towards the FF solutions} \label{sec:towtheffsol}

We present simulations for $\alpha=15^{\circ}$, $\alpha=45^{\circ}$,
and $\alpha=75^{\circ}$. For each of these $\alpha$-values, we have
simulated 11 $\ir$-values that effectively produce an entire
spectrum of solutions from close to VRD to close to FF.
Table~\ref{tab01} shows the adopted $\ir$ values in $\ir_{\rm GJ}^0$
units (second column) and in the corresponding $\ir_{\rm GJ}$ units
(third to fifth column).

Figure~\ref{fig:003} presents, for the indicated $\alpha$-values,
the spin-down power, $\ed$ vs. particle injection rate, $\ir$ in a
log-linear plot. We see that for low and high $\ir$-values $\ed$
asymptotically goes to the values corresponding to VRD and FF ones
denoted by the dashed lines.

\begin{figure*}
\vspace{0.0in}
  \begin{center}
    \includegraphics[width=1.0\linewidth]{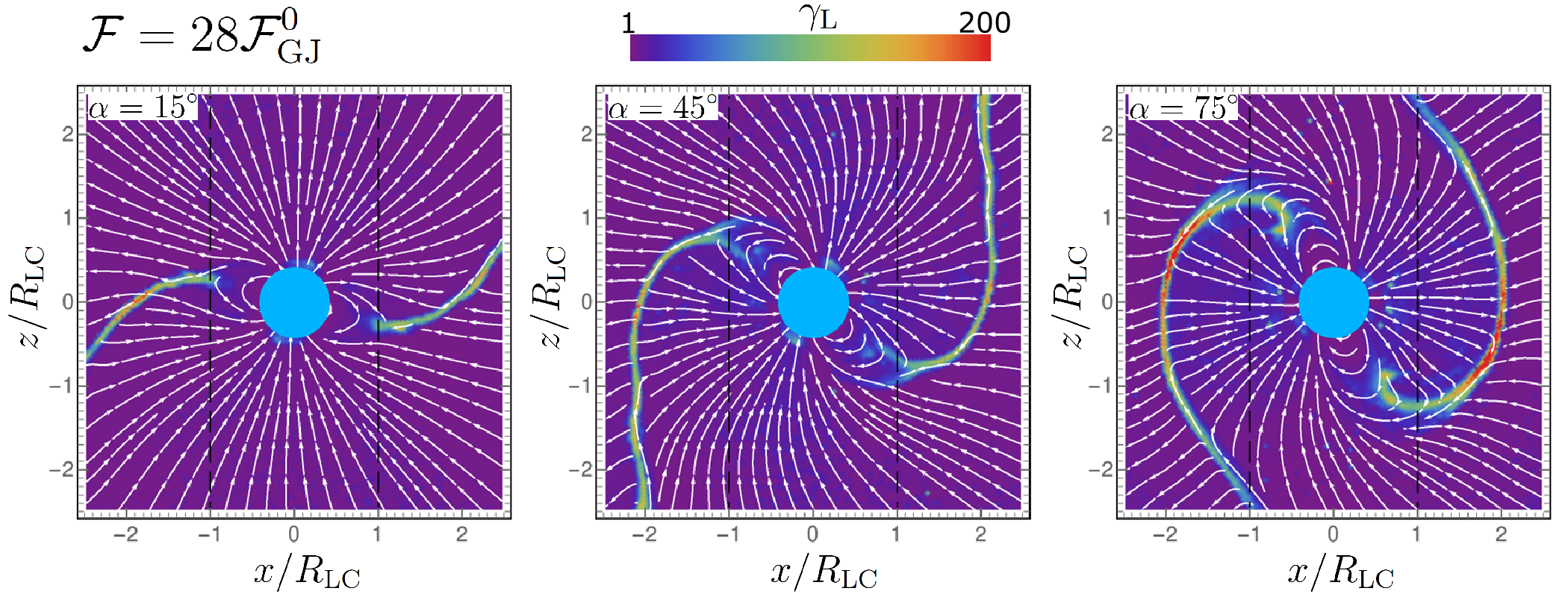}
  \end{center}
  \vspace{0.0in}
  \caption{The average (per computational cell) $\gamma_{\rm L}$
  values in the indicated color scale. The plots are for
  $\ir=28\ir_{\rm GJ}^0$ while each panel corresponds to the indicated
  $\alpha$ value. The high-energy particles lie near the ECS outside the LC
  (black dashed vertical lines).}
  \label{fig:005}
  \vspace{0.0in}
\end{figure*}

In Fig.~\ref{fig:004}, we see the field structure, for
$\alpha=45^{\circ}$ and for the indicated $\ir$-values, on the
poloidal $\pmb{\mu}-\pmb{\Omega}$ plane. The charge density $\rho$
(left-hand column) as well as the number density of positrons
$n_{e^+}$ (middle column) and electrons $n_{e^-}$ (right-hand
column) are shown in the indicated color scales. As $\ir$ increases
features of the FF solution start developing. Thus, the magnetic
field lines open gradually beyond the LC and become straight, while
the ECS beyond the LC starts forming. The low $\ir$ magnetospheres
(i.e. top row) tend to be charge-separated \citep[similar to the
dome-torus configuration;
see~][]{1985MNRAS.213P..43K,2002ASPC..271...81S} while the high
$\ir$-values produce magnetospheres with both kinds of charge
carriers being abundant everywhere (i.e. bottom row).

Figure~\ref{fig:007} shows, in a logarithmic scale on the
$\pmb{\mu}-\pmb{\Omega}$ plane, the number density of the injected
particles per unit time for the indicated $\alpha$-values and for a
high injection rate ($\ir=20\ir_{\rm GJ}^0$). The much higher
densities near the stellar surface are compensated by the much
larger volumes at larger distances. We note that our injection
prescription activates particles injection only when magnetization
$\upsigma_{\rm M}$ exceeds some threshold value
(Eq.~\ref{eq:magnthr}) and so particle injection is strongly
suppressed in the ECS region outside the LC where $B\rightarrow 0$.
This implies that in our simulations the particles appearing
eventually in the ECS originate from the separatrix within the LC
and/or reach the ECS region by drifting across magnetic field lines
(see also the discussion in Section~\ref{sec:fitfermidata}). In
\cite{2017arXiv171003536B}, studying magnetospheres with particles
injected only near the stellar surface, we saw drifting positrons
contributing to the structure of the ECS.

\begin{table}[!tbh]
\centering
        \begin{tabular}{cccc}
            \hline
           & $\alpha=15^{\circ}$ & $\alpha=45^{\circ}$ & $\alpha=75^{\circ}$\\

           $\ir (\ir_{\rm GJ}^0)$ & $\ir (\ir_{\rm GJ})$ & $\ir (\ir_{\rm GJ})$ & $\ir (\ir_{\rm GJ})$\\
             \hline
0.1 & 0.1 & 0.14 & 0.38\\
0.25 & 0.26 & 0.35 & 0.97\\
0.6 & 0.62 & 0.85 & 2.3\\
1.5 & 1.55 & 2.1 & 5.8\\
3.5 & 3.6 & 4.9 & 13.5\\
6.5 & 6.7 & 9.2 & 25.1\\
8.5 & 8.8 & 12.0 & 32.8\\
11.5 & 11.9 & 16.2 & 44.4\\
15.0 & 15.5 & 21.2 & 58.0\\
20.0 & 20.7 & 28.3 & 77.3\\
28.0 & 29.0 & 39.6 & 108.2\\

\hline
        \end{tabular}
    \caption{We present models corresponding to 11 $\ir$ values.
    In the first column, the adopted $\ir$ values are shown
    in the $\ir_{\rm GJ}^0$ unit (i.e. the particle injection rate that
    corresponds to the GJ flux from the two polar caps of the aligned,
    $\alpha=0^{\circ}$, rotator). The second to fourth columns show
    the adopted $\ir$ values in $\ir_{\rm GJ}$ units (i.e. the particle
    injection rate that corresponds to the GJ flux from the two
    polar caps for the indicated $\alpha$ value).}
    \label{tab01}
\end{table}

\begin{figure}
\vspace{0.0in}
  \begin{center}
    \includegraphics[width=0.9\linewidth]{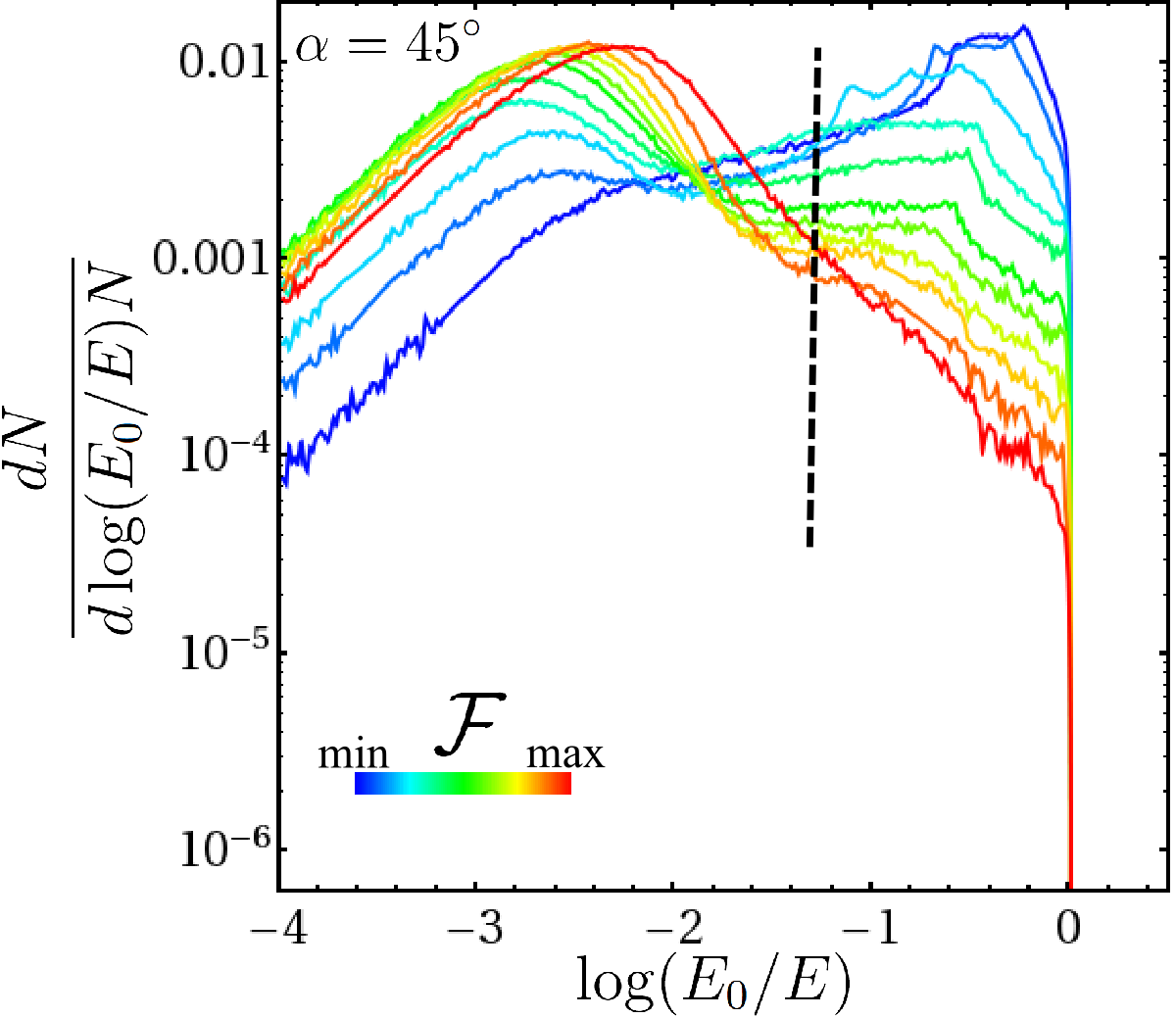}
  \end{center}
  \vspace{0.0in}
  \caption{The distribution, in log-log scale, of the
  $E_0/E$ values calculated at points randomly selected
  within the spherical shell that is defined by the stellar
  surface and the sphere with $r=2.0$. The point density corresponds to
  $\sim 1$ point per computational cell. The plotted results correspond
  to simulations of $\alpha=45^{\circ}$. The different colors correspond
  to different $\ir$ values. As shown in the Figure $\ir$ increases as
  the color changes gradually from blue to red. The vertical dashed line
  makes the value to the right of which we consider that the actual
  acceleration takes place.}
  \label{fig:006}
  \vspace{0.0in}
\end{figure}

Figure~\ref{fig:0045} shows for the indicated $\alpha$, $\ir$ values
the normalized particle energy (i.e. $\gamma_{\rm L}$) distributions
for both $e^{-}$ (blue lines) and $e^{+}$ (red lines). {Even though
the particle energy range, which in reality extends to many orders
of magnitude, is compressed within 3 orders of magnitude, our
simulations are still able to provide not only the current and field
structures corresponding to an entire spectrum of solutions from
vacuum to FF but also the accelerating component of the particle
distribution and the magnetosphere regions where the particle
acceleration takes place. Thus, the particle energy distributions
consist of a low energy population that peaks at $\gamma_{L} \sim 1$
and one at higher energy ($\gamma_{\rm L}\gtrsim 30$) that extends
to $\gamma_L \sim 10^3$.} The first component is the non-accelerated
bulk plasma, which is associated mainly with the short scale ($\sim
\lambda_{\rm D}$) fluctuating fields while the second component is
the accelerating component, which is associated with the larger
scale unscreened fields of the magnetosphere. We note that the
relative strength of the accelerating component decreases with
increasing $\ir$. As we also discuss later a significant part of the
particle acceleration takes place at large distances at and beyond
the LC, even for relatively low $\ir$ (i.e. $\ir\gtrsim 1\ir_{\rm
GJ}^0$). However, for high $\ir$ values and solutions close to the
FF ones the acceleration takes place exclusively at and beyond the
LC in regions near the ECS.

\begin{figure*}[!tbh]
\vspace{0.0in}
  \begin{center}
    \includegraphics[width=0.8\textwidth]{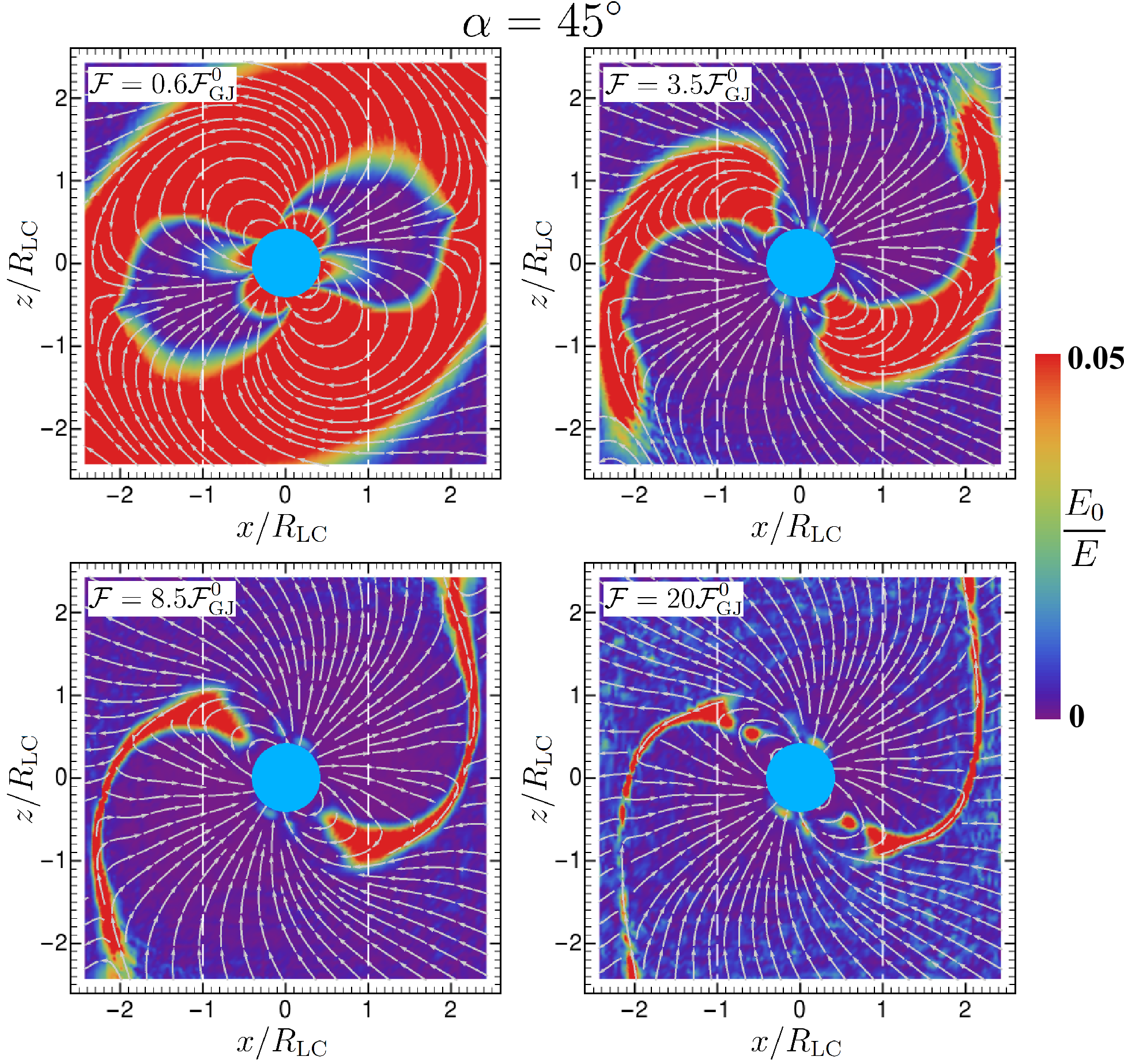}
  \end{center}
  \vspace{0.0in}
  \caption{The spatial distribution of $E_0/E$, in the indicated color
  scale, for simulations of $\alpha=45^{\circ}$. Each panel corresponds
  to the indicated $\ir$ values. The red colored regions correspond
  to the regions above the value indicated by the vertical dashed line
  in Fig.~\ref{fig:006}.}
  \label{fig:008}
  \vspace{0.0in}
\end{figure*}

In Fig.~\ref{fig:005} we plot, for different $\alpha$-values and for
$\ir=28\ir_{\rm GJ}^0$, the average (within a computational cell)
$\gamma_{\rm L}$-value. These simulations have field structures very
close to the FF ones and the average $\gamma_{\rm L}$-values
indicate the existence of energetic particles only in the region
close to the Y-point\footnote{The point where the separatrix meets
the ECS.} and the ECS beyond the LC. The FF ECS for
$\alpha=0^{\circ}$ is positively charged (i.e. $\rho>0$) but as
$\alpha$ increases the corresponding charge density decreases and it
becomes 0 for $\alpha=90^{\circ}$\footnote{At $\alpha=90^{\circ}$
the ECS consists of the same number of $e^+$ and $e^-$.}. This can
be concluded also by the fact that the orthogonal rotator
$\alpha=90^{\circ}$ actually lies in the middle of the aligned and
anti-aligned rotator. In Fig.~\ref{fig:0045} the consistent excess
of positrons at high energies reflects the strong acceleration that
takes place at and beyond the LC near the ECS\footnote{For low $\ir$
values, the notion of the ECS is not unambiguous. However, even in
these cases the high accelerating electric components appear mainly
in the regions that $e^{+}$ dominate and eventually (for high $\ir$
values) will form the ECS.}. The effect is suppressed as we go
towards higher $\alpha$-values.

\section{Interpreting the simulations} \label{sec:rescale}

In the previous section, we saw that an entire spectrum of solutions
from near VRD to near FF ones can be covered by applying arbitrary
particle injection everywhere in the magnetosphere. However, a vital
question that arises is whether and how these simulations can be
compared to the recent rich observational data (i.e. \emph{Fermi}).
The comparison with the observations requires a detailed modeling of
the high-energy emission in PIC simulations. This implies not only
the determination of the emitting regions but also the derivation of
the corresponding $\ec$ and $L_{\gamma}$ values. However, the PIC
particle energies are much smaller than the energies in a real
pulsar magnetosphere (see Fig.~\ref{fig:0045}) due to the much
smaller fields (see Section~\ref{sec:simset}). In addition, there is
no trivial relation (e.g. proportionality) that can provide the
particle energy transformation between the PIC model and the real
system.

\begin{figure*}[!tbh]
\vspace{0.0in}
  \begin{center}
    \includegraphics[width=0.85\textwidth]{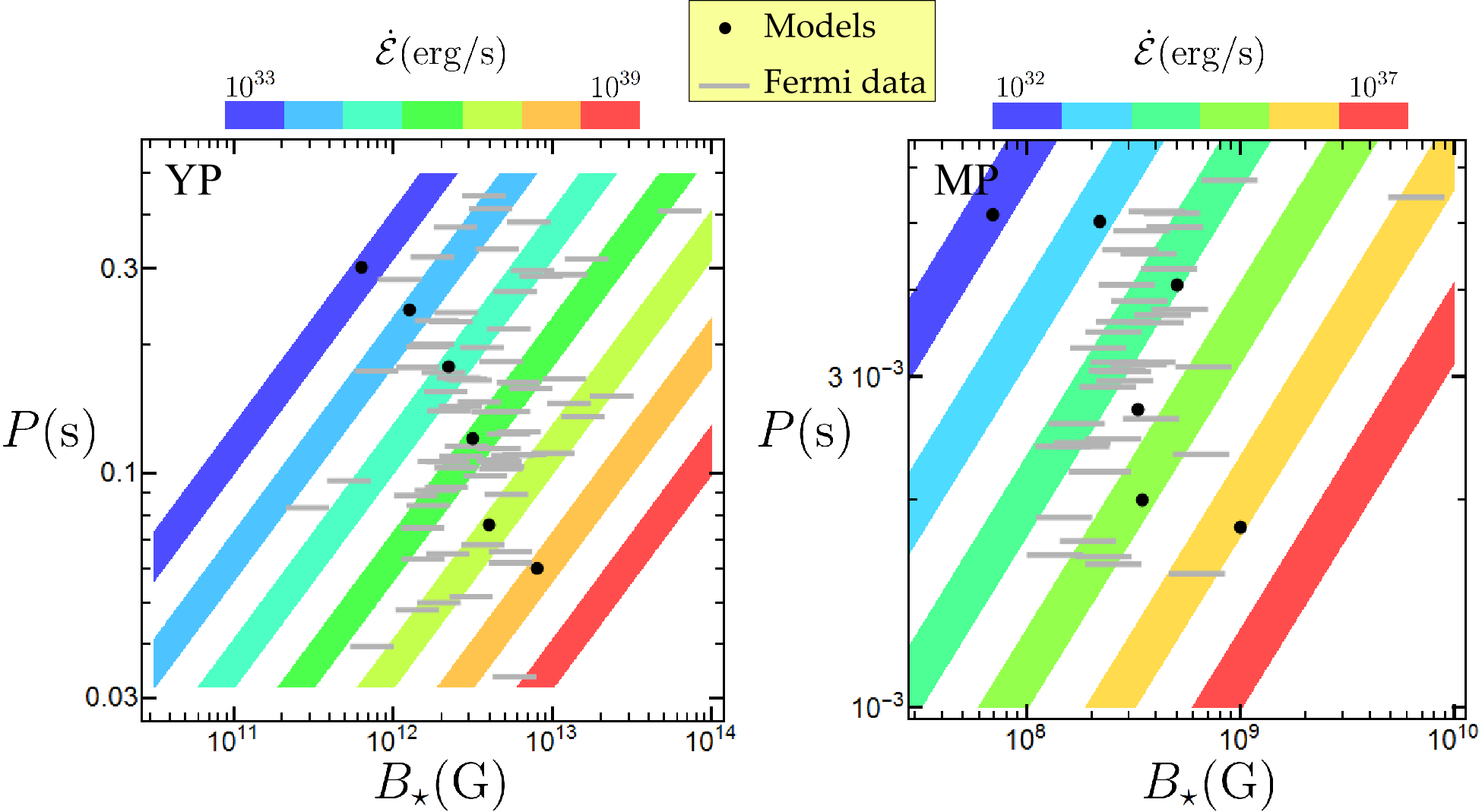}
  \end{center}
  \vspace{0.0in}
  \caption{The $B_{\star}$ vs. $P$ values of the \emph{Fermi} pulsars
  together with the adopted realistic model values (see Table~\ref{tb:realpar}).
  The left and right hand panels correspond to YPs and MPs,
  respectively. The large black dots denote the model points while
  the gray horizontal segments denote the \emph{Fermi} data. The width of
  these segments indicate the $B_{\star}$ uncertainty. More specifically,
  the left and right edges of these segments correspond to
  the $B_{\star}$ value assuming the spin-down power of the
  perpendicular rotator ($\alpha=90^{\circ}$) for the VRD and FF regimes,
  respectively. The colored stripes denote the indicated $\ed$ values with
  the $B_{\star}$ uncertainty discussed above. We see that the 12 model points
  trace well the ``observed" area of the diagrams.}
  \label{fig:009}
  \vspace{0.0in}
\end{figure*}

The particles energies in the real system are not only larger (than
those in the PIC simulations) but are also crucially affected by the
radiation reaction forces with the phenomenon being more prominent
for the high-energy emitting particles. In our PIC simulations, the
impact of the actual radiation reaction forces (Eq.~\ref{eq:frr}) on
the particles energies is {negligible} (due to the corresponding
small $\gamma_{\rm L}$ and $B$ values). Moreover, in principle,
synchrotron and curvature losses act very differently depending
strongly on the specific regime (i.e. $\gamma_{\rm L}$, $B$, radius
of curvature $R_{\rm C}$ of the guiding center). Furthermore, the
gyro-radius in PIC simulations is much larger than in real pulsar
magnetospheres (in $R_{\rm LC}$ units) and in low field regions this
can, in principle, affect magnetosphere features
\citep[see~][]{2015MNRAS.448..606C}.

In principle, the radiation reaction force, $F_{\rm rr}$, on the PIC
particles can be enhanced (artificially) to the level that produces
cooling times equal (in pulsar period units) to those of a realistic
$B_{\star}$ and $P$ magnetosphere values, to establish the correct
loss time scales. However, the realistic cooling times are so small,
especially for the case of high $\ed$ values, that the time steps
needed to capture them are far too short to be implemented in our
simulations. In addition, the maximum particle energies would be
severely reduced relative to the maximum potential drop (because of
the drastic enhancement in $F_{\rm rr}$); this effect would be more
dramatic for the high $\ed$ cases. This reduction would then bring
the entire particle population (including the accelerating
component) either close to or within the statistical noise level
that inherently exists in these simulations, thereby invalidating
any claim of realistic radiation modeling.

In real pulsar magnetospheres, particles will lose their
perpendicular (to the magnetic field) momentum and therefore their
pitch angle very fast due to the corresponding synchrotron losses
(unless there is some mechanism to sustain pitch angles). Under this
regime, particles remain at the zero Landau level thus terminating
the synchrotron radiation emission. Nonetheless, there are, in
principle, certain conditions that can sustain the pitch angles and
make synchrotron emission important. \citet{1998AA...337..433L}
showed that cyclotron resonant absorption of low energy radio
photons is efficient in pulsar magnetosphere environments. In the
regions (at high altitudes within the LC though) where the resonant
condition is fulfilled, the Landau states are populated and the
pitch angles of the particles increase up to the equilibrium point
where the absorption energy gain balances the energy losses due to
(mainly) synchrotron emission. In \cite{2015ApJ...811...63H}
\citep[see also~][]{2008ApJ...680.1378H}, we showed that this
processes, in some cases (e.g. Crab pulsar) may be important for
explaining the optical to hard X-ray observed emission. Moreover,
\cite{2014ApJ...780....3U} showed that in the reconnection region at
the ECS of a Crab-like pulsar equipartition can trigger synchrotron
emission. Nonetheless, any synchrotron emission that is triggered by
cyclotron resonant absorption of low energy radio photons
contributes to energies much smaller than the cutoff energies
observed by \emph{Fermi} while the microphysical emission properties
of the reconnection at the ECS is difficult to be studied properly
in global magnetosphere simulations that resolve rather poorly the
corresponding region.

\begin{figure}[!htb]
\vspace{0.0in}
  \begin{center}
    \includegraphics[width=0.95\linewidth]{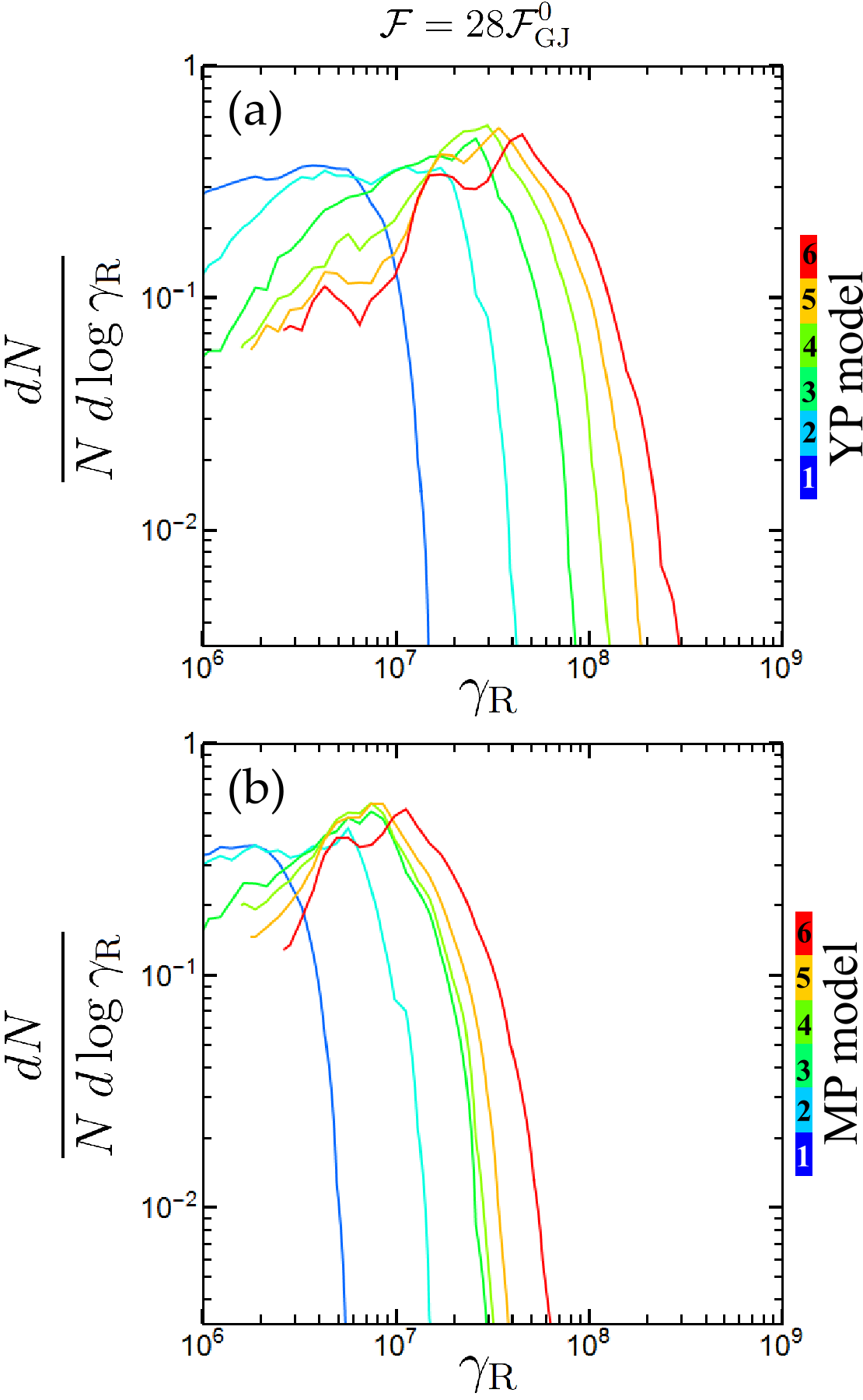}
  \end{center}
  \vspace{0.0in}
  \caption{The distributions of the $\gamma_{\rm R}$ (see Eq.~\ref{eq:gamfict})
  values for $\ir=28\ir_{\rm GJ}^0$ and $\alpha=45^{\circ}$.
  The top and the bottom panels show the
  results for YP and MP models. In each panel the different colors denote the
  indicated model No. (i.e. the different realistic $B_{\star}$, $P$ parameter
  values; see Table~\ref{tb:realpar}).}
  \label{fig:0102}
  \vspace{0.0in}
\end{figure}

Taking into account all the above, we focused on the properties and
efficiency of curvature radiation (CR) assuming that the
contribution of the synchrotron emission is small (especially in the
\emph{Fermi} energy band) because of the rapid decrease of the
perpendicular (to $B$) momentum.

However, the CR calculation depends on both the electric fields the
particles encounter and the orbital radius of curvature. On the one
hand, the electric fields and the corresponding potential drops
depend exclusively on the magnetosphere model (i.e. on the global
particle injection rate). On the other hand, an accurate calculation
of CR implies that the orbital radii of curvature are close to those
corresponding to asymptotic drift approximation trajectories. The
latter ones are actually defined to be those of the so-called
Aristotelian Electrodynamics
\citep[see][]{2013arXiv1303.4094G,2015AJ....149...33K}.

For these asymptotic trajectories, the velocity flow reads
\begin{equation}
\label{eq:ae}
\pmb{\mathrm{v}}=\frac{\mathbf{E}\times\mathbf{B}\pm(B_0\mathbf{B}+E_0\mathbf{E})}{B^2+E_0^2}~.
\end{equation}
where the two signs correspond to the two different types of charge
while the quantities $E_0$ and $B_0$ are related to the Lorentz
invariants \citep{2008arXiv0802.1716G,2012ApJ...746...60L}
\begin{equation}
\label{eq:e0b0} E_0 B_0=\mathbf{E}\cdot
\mathbf{B},\;\;E_0^2-B_0^2=E^2-B^2
\end{equation}
and $E_0$ is the electric field in the frame where $\mathbf{E}$ and
$\mathbf{B}$ are parallel and is the effective accelerating electric
component $\eacc$ which becomes zero only when $\mathbf{E}\cdot
\mathbf{B}=0$ and $E<B$. We note that particles that follow
trajectories defined by Eq.~\eqref{eq:ae} do not emit synchrotron
radiation \citep{2015AJ....149...33K}. For the rest of the paper,
$E_0$ denotes $\eacc$.

In our simulations, the high energy particles, in general, do not
have high pitch angles because on the one hand the particles are
injected with zero velocities and on the other hand the acceleration
along the accelerating electric fields tends to reduce any pitch
angle. Nonetheless, in order to make the CR calculation more
accurate, we used an artificially enhanced $F_{\rm rr}$ in order to
reduce even further the pitch angles. For this, we increased
properly the magnetic fields $B_{\rm eff}$ that enter the radiation
reaction forces on the PIC particles (Eq.~\ref{eq:frr}). The $B_{\rm
eff}$ scaling varies in time (for each particle trajectory) ensuring
always that the corresponding synchrotron cooling time
\begin{equation}
    \label{eq:sct}
    t_{\rm sc}=-\frac{\gamma}{\dot{\gamma}}\propto B_{\rm
    eff}^{-2}\gamma_{\rm L}^{-1}
\end{equation}
is effectively small, which means much smaller than the light
crossing time $t_{\rm l}=1/\Omega$ but always resolved by the
adopted time-step (at least 5 times). {We note that the $B_{\rm
eff}$ fields affect only the radiation reaction forces (i.e.
Eq.~\ref{eq:frr}) and not the ordinary Lorentz force $\mathbf{F_{\rm
L}}=q_e(\mathbf{E}+\mathbf{v}\times\mathbf{B}/c)$ where the actual
(i.e. simulated) fields enter. The resulting trajectories
(especially the high energy ones) have geometric properties closer
to the real ones (i.e. the perpendicular momentum is significantly
reduced). We have actually checked that the vast majority of the PIC
high-energy particles have velocities and orbital radii of curvature
close to those corresponding to AE (i.e. Eq.~\ref{eq:ae}).

\begin{figure*}[!htb]
\vspace{0.0in}
  \begin{center}
    \includegraphics[width=1.0\textwidth]{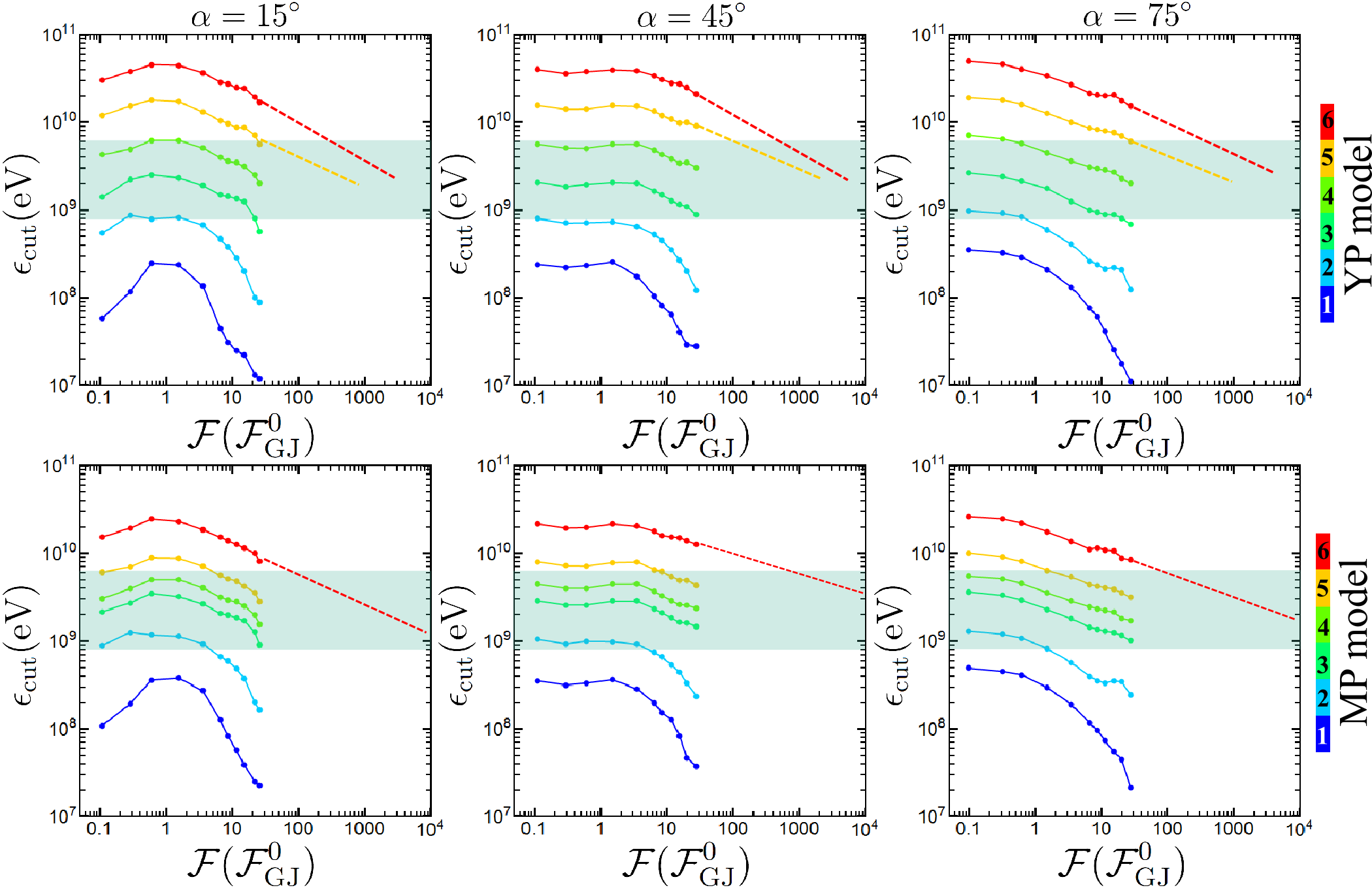}
  \end{center}
  \vspace{0.0in}
  \caption{The $\ec$ values of the model spectra as a function of $\ir$ in
  log-log scale diagrams. \textbf{Left-hand}, \textbf{middle},
  and \textbf{right-hand} columns show
  the results for $\alpha=15^{\circ}$, $\alpha=45^{\circ}$, and
  $\alpha=75^{\circ}$, respectively. The \textbf{top} and \textbf{bottom}
  rows show the results for YP and MP models, respectively.
  In each panel the different colors denote the indicated
  model No. (i.e. the different realistic $B_{\star}$, $P$ parameter
  values; see Table~\ref{tb:realpar}). The light greenish zones
  denote the zone of the observed (by \emph{Fermi}) $\ec$ values
  ($\sim 1-6 \rm GeV$). The dashed lines are extrapolations that provide
  estimations of the $\ir$ values that produce $\ec$ values within the
  \emph{Fermi} (i.e. greenish) zone. {We note that the main
  motivation for the use of linear extrapolations was the apparent
  trend of the decrease of the second derivative of the
  function $\log[\epsilon_{\rm cut}(\log\mathcal{F})]$ for high
  $\mathcal{F}$ values and for high spin-down power models.
  Independently of this, taking into account that the function
  $\log[\epsilon_{\rm cut}(\log\mathcal{F})]$ is convex the
  linear extrapolation provides the highest possible $\mathcal{F}$ value.}}
  \label{fig:010}
  \vspace{0.0in}
\end{figure*}

Nonetheless, the resulting particle energies are still much smaller
than the particle energies in a real pulsar magnetosphere. In order
to derive PIC particle energies of meaningful values, we scale up
the particle energies to the values they would have for real pulsar
surface fields assuming that the trajectories of the high-energy
particle are geometrically correct.} For the calculation of the
scaled-up (i.e. realistic) particle energies, $\gamma_{\rm R}$, done
in parallel inside the PIC simulation, we choose realistic $P,
B_{\star}$-values and we integrate along each particle trajectory
the expression
\begin{equation}
\label{eq:gamfict} \frac{d\gamma_{\rm R}}{dt} =\frac{q_e
\pmb{\mathrm{v}}\cdot \mathbf{E}}{m_e c^2}-\frac{2q_e^2\gamma_{\rm
R}^4}{3R_{\rm C}^2m_e c}
\end{equation}
which provides realistic $\gamma_{\rm R}$ values. We note that in
Eq.\eqref{eq:gamfict}, $R_{\rm C}$ is the local radius of curvature
of the trajectory while $\pmb{\mathrm{v}}$ is the corresponding
particle velocity. {The integration of Eq.~\eqref{eq:gamfict} is
made in parallel with the PIC particle trajectories and it does not
affect the trajectories themselves. It is used in order to provide
realistic estimations (i.e. in real pulsar environments) of particle
energies taking into account the realistic energy gains (i.e. first
term in Eq.~\ref{eq:gamfict}) due to any accelerating electric
fields and the energy losses due to CR reaction losses (i.e. second
term in Eq.~\ref{eq:gamfict}), respectively.} For the calculation of
$R_{\rm C}$, 3 trajectory points are used that are separated in time
$d/c$ where $d$ is the computational cell size. This implies that
the $R_{\rm C}$ value is updated at intervals of $d/c~$\footnote{We
have also checked different intervals (i.e. a few times $d/c$) for
the update of $R_{\rm C}$ and our results remained unaffected.}.
Apparently, as discussed above, $R_{\rm C}$ is not crucially
affected by the (reduced) gyro-motion due to the enhanced
$\mathbf{F_{\rm rr}}$. Thus, in our adopted approach, instead of
using test particles in order to estimate the effectiveness of CR,
we use the particles that the kinetic simulations indicate lie in
the model's strong accelerating regions. Finally, for the
integration of Eq.~\eqref{eq:gamfict}, we take into account only the
effective accelerating electric field components, as described
below.

\begin{figure*}[!tbh]
\vspace{0.0in}
  \begin{center}
    \includegraphics[width=0.8\textwidth]{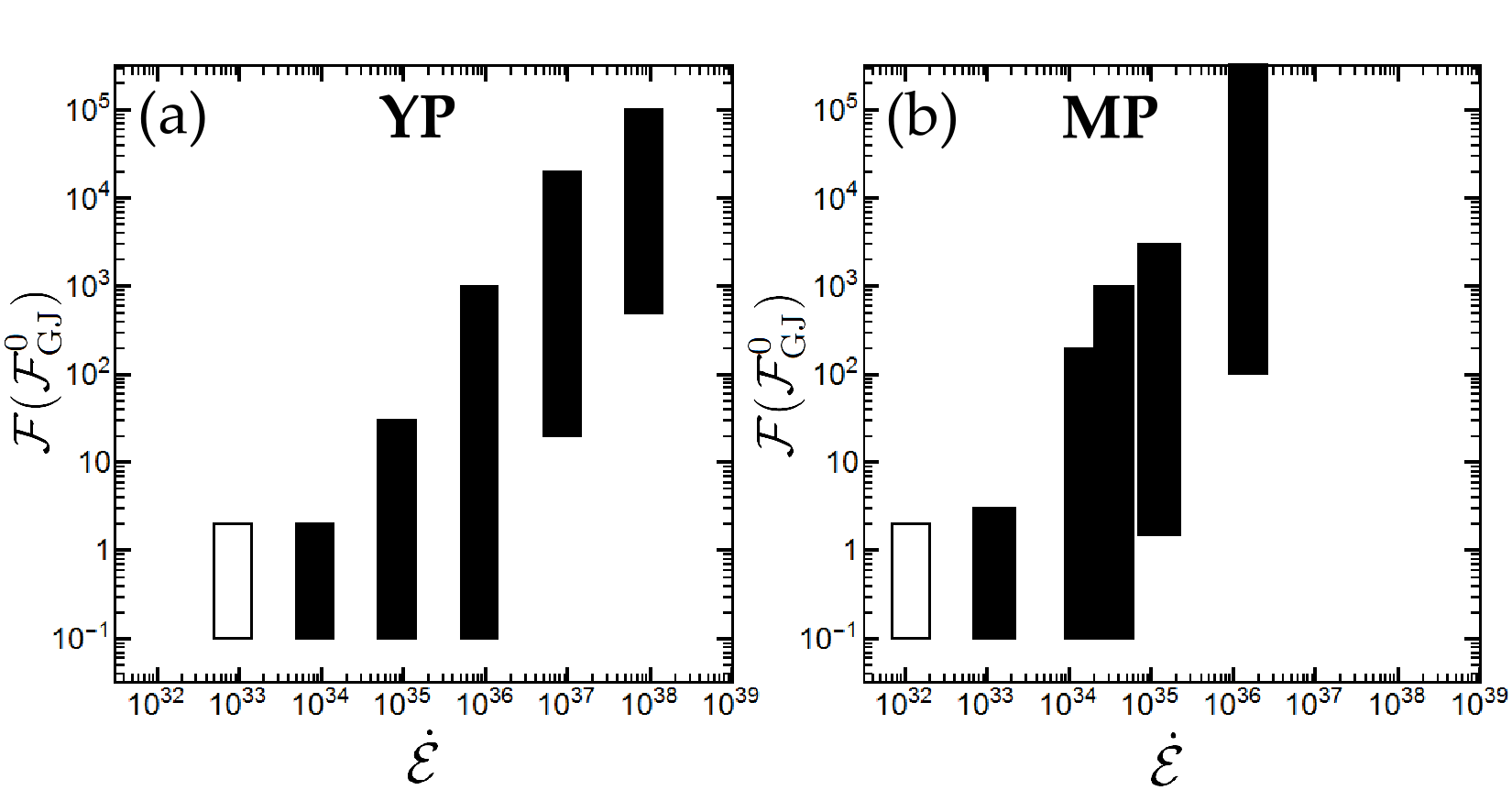}
  \end{center}
  \vspace{0.0in}
  \caption{The $\ir$ value ranges that produce $\ec$ values within the \emph{Fermi}
  zone (see Fig.~\ref{fig:010}) as a function of $\ed$, in log-log scale.
  The rectangle widths denote the $\ed$ range (see Table~\ref{tb:realpar})
  while the rectangle heights denote the optimum $\ir$ range.
  The black filled rectangles indicate models that produce $\ec$ values within
  the \emph{Fermi} zone while the white rectangles indicate models that fail to
  fall within the Fermi $\ec$ values. In the latter case, the corresponding
  $\ir$ range (i.e. vertical dimension) denotes the $\ir$ values that produce
  $\ec$ close to the \emph{Fermi} zone. We see that for high $\ed$ YPs and MPs the
  optimum $\ir$ can reach up to $~10^5$.}
  \label{fig:0105}
  \vspace{0.0in}
\end{figure*}

In our PIC simulations there are magnetosphere regions where nearly
all the corresponding accelerating electric components are screened
and magnetosphere regions with consistent accelerating electric
components. The former regions are the effective FF regions where
the field fluctuations dominate while the latter regions are the
actual accelerating locations. We have found that the best way to
identify these regions is to use the local values of the normalized
effective accelerating electric field $E_{0}/E$, where $E$ is the
local total electric field {(in the observer's inertial frame)}. In
Fig.~\ref{fig:006}, we plot, in $\log-\log$ scale, the distributions
of $E_{0}/E$ values that have been calculated at points that are
uniformly distributed within the magnetosphere volume that extends
from the stellar surface up to $r=2.5R_{\rm LC}$. These results are
for simulations with $\alpha=45^{\circ}$ corresponding to the
different particle injection rates ($\mathcal{F}$; see
Table~\ref{tab01}) that are denoted by the different colors
according to the indicated color scale. We observe the presence of a
maximum at $\log(E_0/E)\lesssim -2$ associated with a thermal-like
particle distribution along with a bump that always appears in these
distributions at higher values. The thermal-like maximum is related
to the fluctuating (noisy) field behavior; its value depends
primarily on the time-step and the number of macroparticles in our
simulations. The higher value component (i.e. bump), beyond the
vertical dashed line, actually indicates the magnetosphere regions
where the true high acceleration takes place. We have checked that
the $E_{0}/E$-position of the maximum decreases (increases) as we
reduce (strengthen) noise. Thus, when we increase the number of
macro-particles and/or decrease the time-step and/or increase the
spatial resolution and/or increase the effective area of the current
smoothing kernel the maximum of the distributions shown in
Fig.~\ref{fig:006} moves to lower $\log(E_0/E)$ values. In
Fig.~\ref{fig:006}, the gradual drift of the maximum towards higher
$\log(E_0/E)$ as $\mathcal{F}$ increases is due to the corresponding
poorer resolution for $\lambda_{\rm D}$ and $\omega_{\rm p}$ (aka
higher noise level). The increase of $\mathcal{F}$ implies higher
particle number densities and consequently smaller $\lambda_{\rm D}$
and higher $\omega_{\rm p}$. Keeping all the other parameters (e.g.
$d$, $dt$) the same, these lower $\lambda_{\rm D}$ and higher
$\omega_{\rm p}$ values are more poorly resolved. This gradually
enhances the noise leading to the increase of the position of the
maximum. On the other hand, we see that the area under the
non-thermal extension (i.e. bump) decreases with increasing
$\mathcal{F}$. This implies smaller magnetosphere volumes with
consistent accelerating electric field components and lower actual
accelerating electric fields. Actually, Fig.~\ref{fig:006} indicates
also our limitations for the highest $\mathcal{F}$ values we can use
for a specific parameter set. Thus, as $\mathcal{F}$ increases and
the $\log(E_0/E)$ maximum moves towards higher values, the gradually
reducing area under the bump eventually (at some $\mathcal{F}$
value) is buried under the corresponding noise.

In Fig.~\ref{fig:008} we show, in the indicated linear color scale,
the $\log(E_0/E)$ values on the poloidal $\pmb{\mu}-\pmb{\Omega}$
plane for $\alpha=45^{\circ}$ simulations and for the indicated
$\mathcal{F}$ values. The reddish regions denote the regions with
values higher than the value corresponding to the dashed vertical
line shown in Fig.~\ref{fig:006}. We see that at lower $\mathcal{F}$
values large magnetosphere regions have consistent accelerating
electric fields while for higher $\mathcal{F}$ values the reddish
regions shrink and finally are restricted to only near the ECS
outside the LC. For all the model results presented below, we assume
that the particle acceleration that produces the high-energy
emission takes place in the magnetosphere regions corresponding to
values greater than the one indicated by the dashed line in
Fig.~\ref{fig:006}. We have decided to use the same threshold value
for consistency even though the bump appears at slightly different
values. However, we have checked that taking into account the
regions corresponding to smaller $\log(E_0/E)$ values (for the
simulations where the bump appears earlier than the dashed line in
Fig.~\ref{fig:006}) do not change the results, mainly because the
corresponding accelerating electric components are small. Moreover,
we have seen that the ohmic $\pmb{\mathrm{J}}\cdot \pmb{\mathrm{E}}$
dissipation takes place largely in the magnetosphere regions
corresponding to $\log(E_0/E)$ values higher than the dashed line
threshold. For the rest of the magnetosphere the total
$\pmb{\mathrm{J}}\cdot \pmb{\mathrm{E}}$ dissipation tends to zero.

Finally, we note that in real pulsar magnetospheres, the noise level
(i.e. the maximum of the distribution in Fig.~\ref{fig:006}) is
expected to be much smaller relative to the maximum of $E_0/E$ in
our simulations. This allows, in principle, the existence of
consistent accelerating electric fields for much lower $\log(E_0/E)$
values than those we assume in our study. Nonetheless, these
accelerating electric fields are expected to be important for the
lower part of the emission spectrum (and not for energies near the
$\ec$ values observed by \emph{Fermi}).

\begin{figure*}[!tbh]
\vspace{0.0in}
  \begin{center}
    \includegraphics[width=0.85\textwidth]{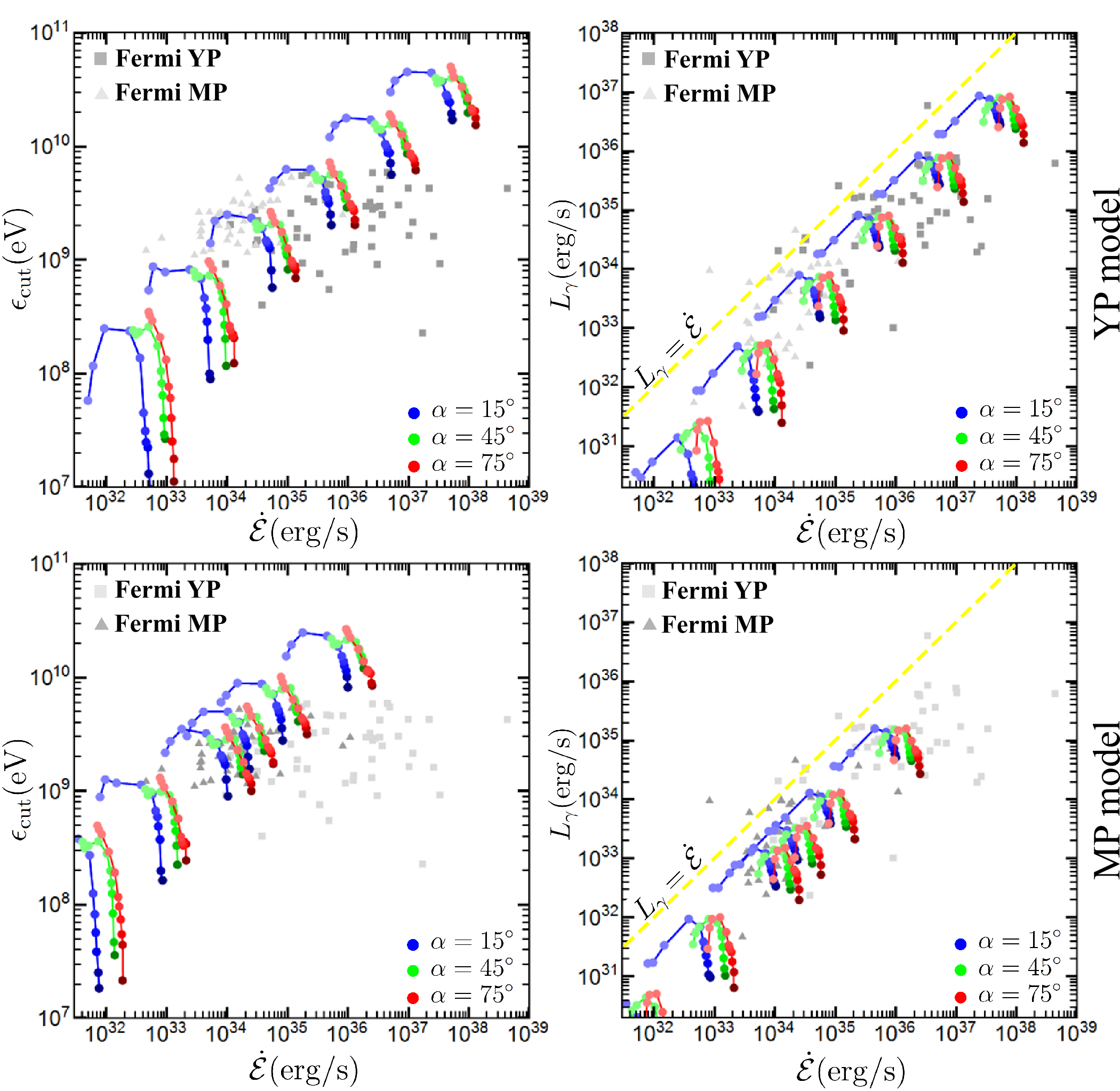}
  \end{center}
  \vspace{0.0in}
  \caption{The model $\ec$ values (left-hand column) and the model $L_{\gamma}$
  values (right-hand column) together with the corresponding \emph{Fermi} ones (2PC)
  as a function of $\ed$ for YPs (top row) and MPs (bottom row). The
  blue, green, and red color denote simulations for $\alpha=15^{\circ}$,
  $\alpha=45^{\circ}$, and $\alpha=75^{\circ}$, respectively. Each
  line-connected set of points corresponds to the same $B_{\star}$ and $P$ values
  (see Table~\ref{tb:realpar}) while the darker the color hue the higher
  the corresponding $\ir$ value is. The \emph{Fermi} data are denoted by gray
  rectangles (YPs) and triangles (MPs). The dashed yellow lines on the
  right-hand panels indicate 100\% $\gamma$-ray efficiency (i.e. $L_{\gamma}=\ed$).
  The optimum $\ir$ ranges that produce
  $\ec$ within the \emph{Fermi} zone (see Figs.~\ref{fig:010} and \ref{fig:0105})
  reproduce the observed $L_{\gamma}$ behavior (i.e. $L_{\gamma}\propto \ed$
  for low $\ed$ and $L_{\gamma}\propto \sqrt{\ed}$ for high $\ed$). Our models
  imply that the observed $L_{\gamma}$ dispersion could be the result of
  different $\ir$ values, different $\alpha$ values, and a variation of the
  beaming-factor, $F_{\rm b}$ with the observer angle, $\zeta$.}
  \label{fig:011}
  \vspace{0.0in}
\end{figure*}

\section{Fitting the \emph{Fermi} data} \label{sec:fitfermidata}

In real pulsars, the direct measurements of the period $P$ and its
derivative $\dot{P}$ allow the calculation of the spin-down power.
Assuming that the moment of inertia $I$ is known the spin-down power
is given by
\begin{equation}
\label{eq:sdpI} \ed=\frac{4\pi^2 I \dot{P}}{P^3}
\end{equation}
which can indirectly lead to an estimation of the magnetic field on
the stellar surface. However, the spin-down power depends on the
magnetosphere regime and is given by \citep{deutsch1955,S2006}
\begin{subequations}
\begin{mathletters}
\label{eq:pfvmff}
\begin{eqnarray}
    \ed=\frac{8\pi^4 r_{\star}^6}{3c^3}\frac{B_{\star}^2}{P^4}\sin^2\alpha
    \;\;\;\text{for the VRD regime,}\phantom{\hspace{29pt}}\\
    \ed=\frac{4\pi^4
    r_{\star}^6}{c^3}\frac{B_{\star}^2}{P^4}(1+\sin^2\alpha)
    \;\;\;\text{for the FF regime.}\phantom{\hspace{29pt}}
\end{eqnarray}
\end{mathletters}
\end{subequations}
By combining Eqs.\eqref{eq:sdpI},~\eqref{eq:pfvmff} we can get the
$B_{\star}$ value with an uncertainty due to the unknown regime and
the unknown $\alpha$ value. In the left (right) hand panel of
Fig.~\ref{fig:009}, we plot on $(P,~B_{\star})$ diagrams all the
\emph{Fermi} YPs (MPs). Each pulsar is represented by a horizontal
light gray line that denotes the uncertainty in $B_{\star}$
estimation. We note that the left and right ends of the light gray
lines denote the $B_{\star}$ values that correspond to the
$\alpha=90^{\circ}$ at the FF and VRD regimes, respectively. The
color stripes denote the indicated $\ed$ values using the same
$B_{\star}$ uncertainty as the light gray lines for the \emph{Fermi}
pulsars.

For the study of the high-energy emission in our simulations, we
follow the approach that is discussed in the previous section.
Table~\ref{tb:realpar} shows the realistic $P,~B_{\star}$ values
that we have chosen. The calculations described in the previous
section are computationally demanding and therefore we have selected
only 12 cases (6 for YPs and 6 for MPs) that trace well the
corresponding observed area (see the black points in
Fig.~\ref{fig:009}).

\begin{figure*}[!tbh]
\vspace{0.0in}
  \begin{center}
    \includegraphics[width=0.85\textwidth]{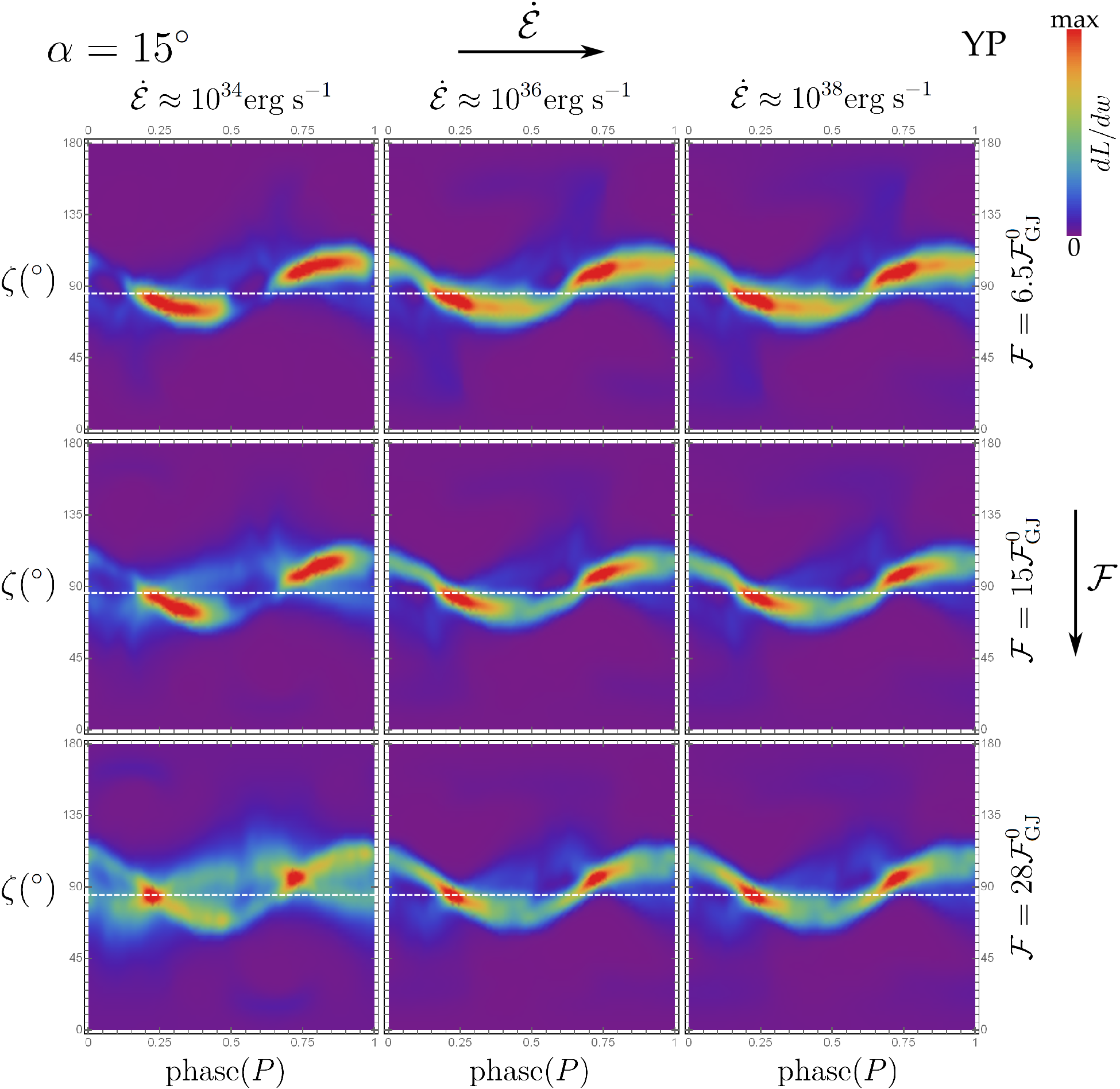}
  \end{center}
  \vspace{0.0in}
  \caption{The sky-maps (i.e. luminosity per solid angle) for
  $\alpha=15^{\circ}$ presented in the indicated linear color scale.
  Each column (row) corresponds to the indicated $\ed$ ($\ir$)
  values. The white horizontal dashed lines indicate the $\zeta$
  value that correspond to the light-curves shown in Fig.~\ref{fig:0141}.}
  \label{fig:012}
  \vspace{0.0in}
\end{figure*}

By calculating the realistic particle $\gamma_{\rm R}$ values, we
can calculate not only the individual particle emissivity ($\propto
\gamma_{\rm R}^4 R_{\rm C}^{-2}$; see Eq.~\ref{eq:gamfict}) but also
the entire CR spectrum of the corresponding emission. In
Fig.~\ref{fig:0102}, we plot, in $\log-\log$ scale, the normalized
distribution of the realistic $\gamma_{\rm R}$ values corresponding
to a simulation with $\alpha=45^{\circ}$ and $\mathcal{F}=28F_{\rm
GJ}^0$. The top and bottom panels show the distributions for the
different $P,~B_{\star}$ values that correspond to YPs and MPs,
respectively.

Taking into account the emission from the entire magnetosphere (up
to $2.5R_{\rm LC}$) we construct the ``global" spectrum, which we
then fit with the model used in 2PC, namely,
\begin{equation}
    \label{eq:spfit}
    \frac{dN}{d\epsilon}=A \epsilon^{-\Gamma}\exp\left(-\frac{\epsilon}{\epsilon_{\rm
    cut}}\right)
    \vspace{-0.05in}
\end{equation}
where $\Gamma$ is the photon-index.

\begin{figure*}[!tbh]
\vspace{0.0in}
  \begin{center}
    \includegraphics[width=0.85\textwidth]{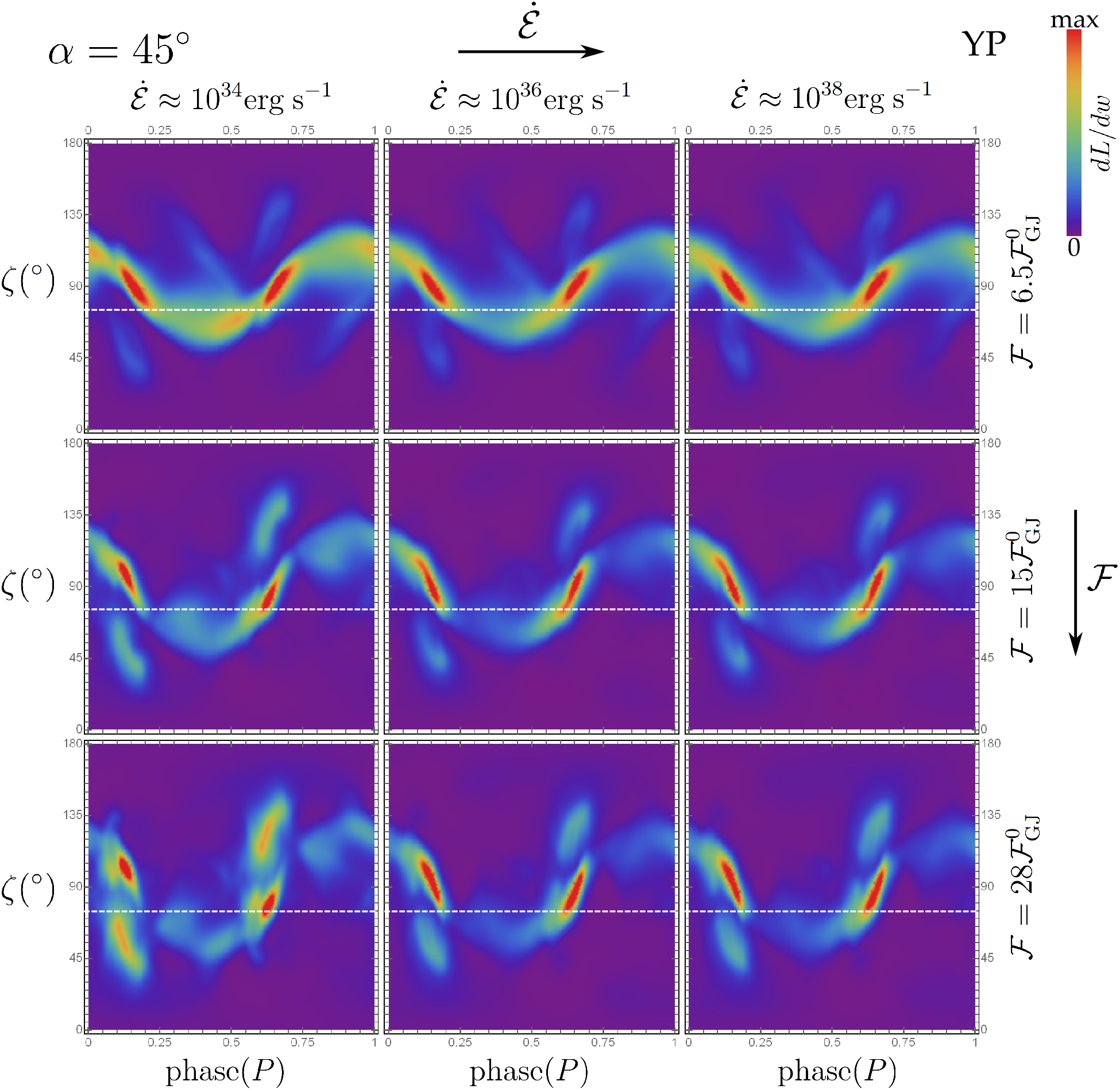}
  \end{center}
  \vspace{0.0in}
  \caption{Similar to Fig.~\ref{fig:012} but for $\alpha=45^{\circ}$.
  The corresponding light-curves are shown in Fig.~\ref{fig:0142}.}
  \label{fig:013}
  \vspace{0.0in}
\end{figure*}

Figure~\ref{fig:010} shows, in $\log-\log$ scale, the model $\ec$
values as a function of the particle injection rates $\mathcal{F}$.
The three different columns show the results for the indicated
$\alpha$ values while the top and bottom rows show the results for
the YP and MP model cases, respectively. The light greenish zones in
each panel of Fig.~\ref{fig:010} denote the zone of the observed (by
\emph{Fermi}) $\ec$ values ($\sim 1-6 \rm GeV$)\footnote{{For a
minority of all the 2PC pulsars ($\lesssim 20\%$) a subexponential
fit model is statistically preferred over the exponential one. The
corresponding $\ec$ can then be up to an order of magnititude
smaller. Nonetheless, the apex energies (of the
$\epsilon^2dN/d\epsilon$ spectrum) for most of these objects are
around a GeV.}}. The different colors indicate the results for the
different $P,~B_{\star}$ (i.e. $\ed$) values that are shown in
Table~\ref{tb:realpar}. First, we see that, for the same
$\mathcal{F}$ value, the higher the $\ed$ the higher the
corresponding $\ec$ value. This increase is due to the corresponding
higher accelerating electric components $E_{0}$. We see also that
for the same $\ed$ value (i.e. along a line) and for the high
$\mathcal{F}$ values $(\gtrsim1\mathcal{F_{\rm GJ}^0})$, $\ec$
decrease with $\mathcal{F}$. This decrease is more prominent for low
$\alpha$ and $\ed$ values. On the other hand, towards low
$\mathcal{F}$ values, $\ec$ either decreases (for low $\alpha$) or
tends to stabilize (for high $\alpha$). We have already discussed
that low $\mathcal{F}$ values produce solutions near VRD whose
spin-down power (i.e. Poynting Flux), for low $\alpha$ values goes
to very low values (see Eq.~\ref{eq:pfvmff}a). The low $\ed$ values
imply a reservoir of low $E$ fields and therefore low accelerating
electric components $E_{0}$ ($\leq E$). This is mainly the reason
why, for low $\alpha$ values, $\ec$ values decrease towards low
$\mathcal{F}$.

Figure~\ref{fig:010} provides a possible explanation for why we
start observing YPs and MPs for spin-down powers $\gtrsim 10^{34}\rm
erg\;s^{-1}$ and $\gtrsim 10^{33}\rm erg\;s^{-1}$, respectively. We
see that for low $\mathcal{F}$ values the low $\ed$ YP and MP models
(see Table~\ref{tb:realpar}) struggle to reach the zone observed by
\emph{Fermi} ($\approx 1\rm GeV$). Taking also into account now the
\emph{Fermi} threshold ($\approx 100\rm MeV$) and the sensitivity of
the instrument, we understand that YPs and MPs with $\ed$ values
lower than those observed are difficult to be detected by
\emph{Fermi}. These objects, if they exist, should have spectra with
$\ec$ values lower than those observed by \emph{Fermi} (at the MeV
levels). We note here that all the MeV pulsars detected so far
\citep{2015MNRAS.449.3827K} have high $\ed$ values. {However, the
light-curve characteristics of these objects indicate that the
corresponding line-of-sight (i.e. $\zeta$ value) might not cross the
core of the high-energy emission, particularly for small $\alpha$
and for intermediate $\zeta$ values \citep[see Fig.~\ref{fig:012};
see also][]{2017arXiv171202406H}.}

On the other hand, models with higher $\ed$ values lie within the
\emph{Fermi} zone for a range of $\ir$ values adopted in the current
study. We see also that the high $\ed$ models ($\gtrsim 10^{37}\rm
erg~s^{-1}$ for YPs and $\gtrsim 10^{35}\rm erg~s^{-1}$ for MPs)
require higher $\mathcal{F}$ values than those we have applied.
However, higher $\mathcal{F}$ values are a computationally very
demanding task and so we get an estimation of the corresponding
$\mathcal{F}$ values by linear extrapolations (see dashed lines in
Fig.~\ref{fig:010}).

\begin{figure*}[!tbh]
\vspace{0.0in}
  \begin{center}
    \includegraphics[width=0.85\textwidth]{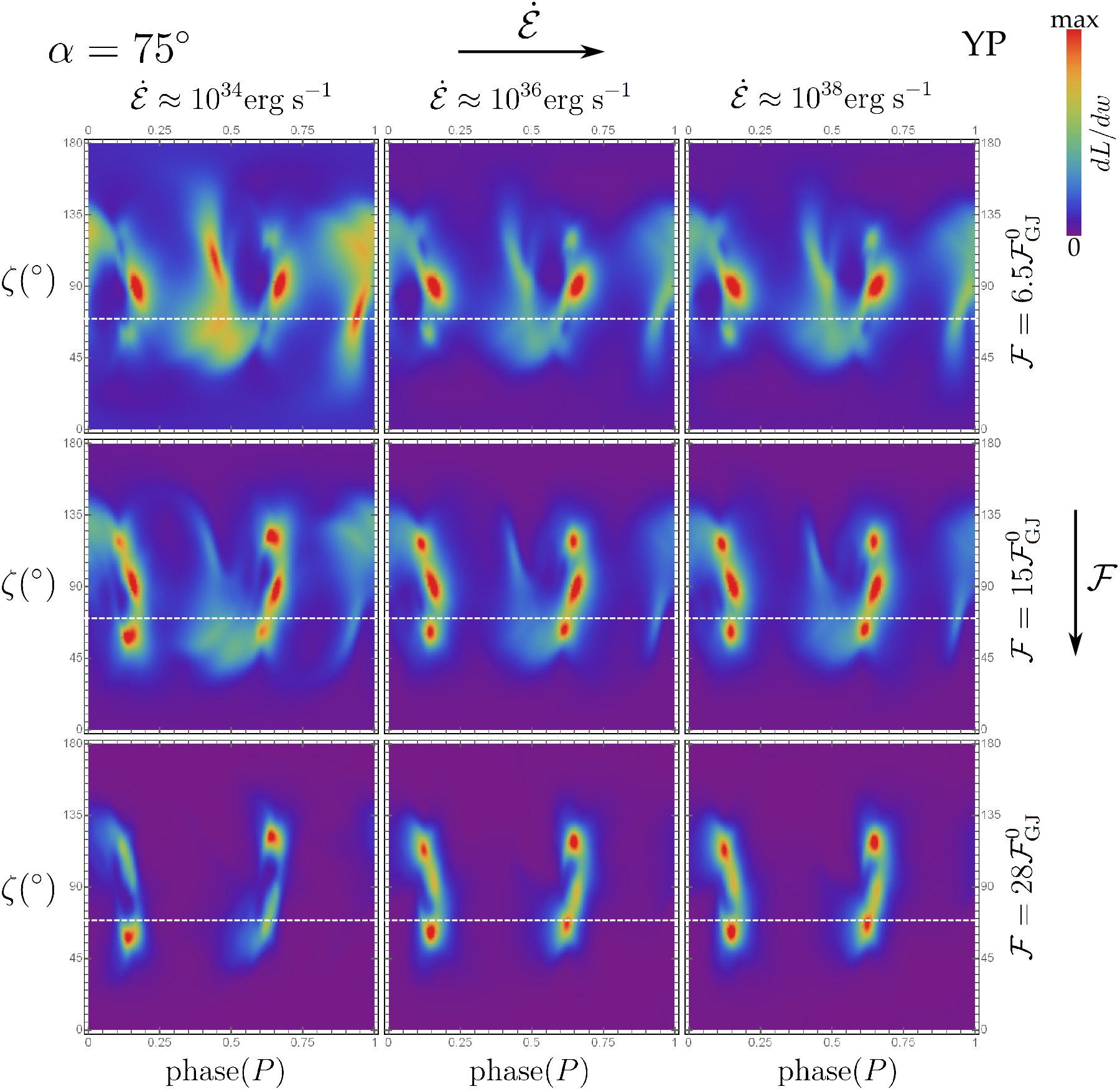}
  \end{center}
  \vspace{0.0in}
  \caption{Similar to Fig.~\ref{fig:012} but for $\alpha=75^{\circ}$.
  The corresponding light-curves are shown in Fig.~\ref{fig:0143}.}
  \label{fig:014}
  \vspace{0.0in}
\end{figure*}

In Fig.~\ref{fig:0105}, we plot, in $\log-\log$ scale, the
$\mathcal{F}$ value as a function of $\ed$ that produces $\ec$
values within the \emph{Fermi} zone . Left (right) hand panel shows
the results for YP (MP) models. The horizontal dimension of each
rectangle denotes the model $\ed$ range (see Table~\ref{tb:realpar})
while the vertical dimension denotes the model $\mathcal{F}$ range
(for all $\alpha$ values) that provide $\ec$ values within the
\emph{Fermi} zone. We note that for the optimum $\mathcal{F}$ range
we have taken into account the extrapolations shown in
Fig.~\ref{fig:010}. The empty rectangles denote that no
$\mathcal{F}$ value can take the models within the \emph{Fermi}
zone. In this case the vertical $\mathcal{F}$ range denotes the
values that produce the closest to the \emph{Fermi} zone $\ec$
values. We see that the highest $\mathcal{F}$ for the high $\ed$
models can reach up to $10^5-10^6 \ir_{\rm GJ}^0$ which is
consistent with the particle multiplicities found in local pair
cascades studies \citep{2015ApJ...810..144T} and to the
multiplicities that are needed for the explanation of the Crab
nebula spectrum \citep{1996ApJ...457..253D}.

\begin{figure*}[!tbh]
\vspace{0.0in}
  \begin{center}
    \includegraphics[width=1.0\textwidth]{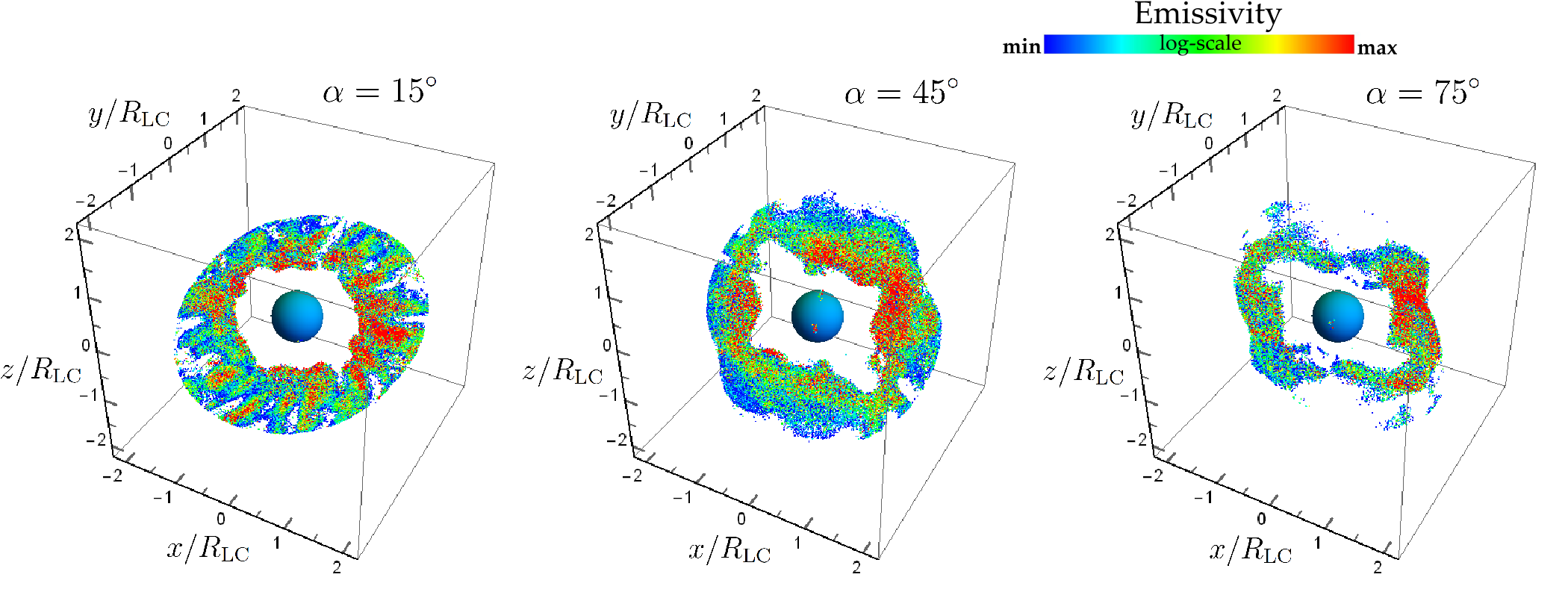}
  \end{center}
  \vspace{0.0in}
  \caption{A random sample of the particles in the 3D space
  that produce the highest 95\% of the observed radiated emission. The colors denote
  the emissivity according to the indicated logarithmic color scale.
  The max value indicated in the color scale is half an order of magnitude
  higher than the corresponding min one. The emission, for high $\alpha$ values,
  is not uniformly distributed in the ECS locus but it is more concentrated
  around the rotational equator. This effect had been originally seen in the
  macroscopic dissipative models \citep{kalap2014}.}
  \label{fig:015}
  \vspace{0.0in}
\end{figure*}
\begin{table*}[!hbt]
\centering
        \begin{tabular}{ccccccc}
            \hline
\multicolumn{1}{c|}{Model} & \multicolumn{3}{c|}{Young Pulsars} & \multicolumn{3}{c}{Millisecond Pulsars}\\
            \cline{1-7}\\[-8pt]
     No.     &  $B_{\star}$ & $P$ & $\dot{\mathcal{E}}_{\rm FF}$ & $B_{\star}$ & $P$ & $\dot{\mathcal{E}}_{\rm FF}$ \\
            &  ($10^{12}$G) & (ms) & ($\rm erg\;s^{-1}$) & ($10^{8}$G) & (ms) & ($\rm erg\;s^{-1}$) \\ \hline
1 & 0.63 & 302.0 & $(0.5-1.4)\times 10^{33}$ & 0.7 & 5.1 & $(0.7-2.0)\times 10^{32}$\\
2 & 1.26 & 239.9 & $(0.5-1.4)\times 10^{34}$ & 2.2 & 5.0 & $(0.7-2.2)\times 10^{33}$\\
3 & 2.24 & 177.8 & $(0.5-1.4)\times 10^{35}$ & 5.0 & 4.1 & $(0.9-2.6)\times 10^{34}$\\
4 & 3.16 & 120.2 & $(0.5-1.4)\times 10^{36}$ & 3.3 & 2.7 & $(2.0-6.0)\times 10^{34}$\\
5 & 3.98 & 75.9 & $(0.5-1.4)\times 10^{37}$ & 3.5 & 2.0 & $(0.7-2.2)\times 10^{35}$\\
6 & 7.94 & 60.3 & $(0.5-1.4)\times 10^{38}$ & 10.0 & 1.8 & $(0.9-2.6)\times 10^{36}$\\
\hline
        \end{tabular}
    \caption{The 12 (6+6) adopted realistic ($B_{\star},~P$) value sets for YP and MP
    models. For each value set the FF spin-down power range is indicated
    (see Eq.~\ref{eq:pfvmff}b).}
    \label{tb:realpar}
\end{table*}

In Fig.~\ref{fig:011}, we plot the $\ec$ (left-hand column) and
$L_{\gamma}$ (right-hand column) values for all the YP (top row) and
MP (bottom row) models together with the \emph{Fermi} data, as
functions of the corresponding spin-down power, $\ed$. We note that
the darker the model points the higher the corresponding
$\mathcal{F}$ value. Similarly to Fig.\ref{fig:010},
Fig.~\ref{fig:011} reveals (from a different perspective than
Fig.~\ref{fig:010}) the model $\mathcal{F}$ value ranges that are
consistent with both $\ec$ and $L_{\gamma}$ \emph{Fermi} values. In
the right-hand column, we see that a maximum appears as the
$\mathcal{F}$ varies from low to high values. This expected behavior
reflects the ideal nature of the VRD and FF solutions. The former
solutions have high fields but limited numbers of particles to
dissipate the energy while the latter ones have high numbers of
particles but limited energy available for dissipation. Taking into
account the corresponding $\ed$ value, we see that the maximum
efficiency ranges from $\approx 13\%$ to $\approx 35\%$ and is
higher (lower) for the lower (higher) $\alpha$ values. We note that
Fig.~\ref{fig:011} implies that the optimum $\mathcal{F}$ values
that reproduce the 2PC $L_{\gamma}$ values are somehow higher then
the optimum $\mathcal{F}$ values that reproduce the 2PC $\ec$
values. Nonetheless, as we have discussed in
\citep{2017ApJ...842...80K}, the 2PC $L_{\gamma}$ values are less
reliable than the corresponding $\ec$ values. The former ones depend
on the assumed beaming-factor $F_{\rm b}$ and the estimated
distances while their wide spread with $\ed$ indicates that other
factors (i.e. $\alpha$-values, variability of $F_b$ with
observer-angle, $\zeta$) play an important role on their
determination. On the other hand, the range of $\ec$ values is more
limited, it does not suffer from geometry or distance uncertainties,
and depends weakly only on the adopted fit-model. A detailed study
of the model $F_b$ values that takes into account the
observationally dependent parameters (e.g. $\alpha$, $\zeta$ values,
distances, instrument thresholds etc.) could eventually answer
whether the observed $\gamma$-ray fluxes are consistent either with
the model $L_{\gamma}$ values corresponding to the optimum
$\mathcal{F}$ values that reproduce the observed $\ec$ values or to
the 2PC $L_{\gamma}$ values, which assume always $F_b=1$. Even
though this study goes beyond the scope of the present paper, our
preliminary calculations indicate that the average model $F_b$
values are smaller than 1, which means, assuming the model
correctness, that the 2PC $L_{\gamma}$ values are overestimated.
Finally, we note that the $\ec$ of MPs for the same $\ed$ and $\ir$
are $\sim 3\times$ higher than the corresponding ones of YPs, which
is consistent with the \emph{Fermi} data
\citep{2017ApJ...842...80K}. This results from the $\ec\propto
B_{\star}^{-1/8}$ relation that is presented in
\cite{2017ApJ...842...80K} and emanates from the assumption that the
emission takes place at the LC near the ECS due to CR at the
radiation-reaction-limit-regime. Taking into account that
${B_{\star_{\rm MP}}}\approx 10^{-4}{B_{\star_{\rm YP}}}$ the
previous relation leads to ${\epsilon_{\rm cut_{\rm
MP}}}\approx{3\epsilon_{\rm cut_{\rm YP}}}$.

In Fig.~\ref{fig:012}, \ref{fig:013}, and \ref{fig:014}, we plot
sky-map atlases for $\alpha=15^{\circ}$, $\alpha=45^{\circ}$, and
$\alpha=75^{\circ}$, respectively. In each figure, the different
rows correspond to the different indicated particle injection rates
($\mathcal{F}$) while the different columns correspond to the
indicated different YP models (i.e. different $P,~B_{\star}$ values
and therefore different $\ed$ values; see Table~\ref{tb:realpar}).
Each sky-map shows, in the indicated color scale, the emitted
luminosity per solid angle ($dL/dw=dL/\sin(\zeta)d\zeta d\phi_{\rm
ph}$). The horizontal axis depicts the phase, $\phi_{\rm ph}$ of the
pulsar rotation while the vertical axis depicts the observer
inclination angle, $\zeta$, which is the angle between the
rotational axis and the line of sight. Horizontal cuts of each
sky-map correspond to the light-curve that is observed by an
observer that lies at some $\zeta$ value. We note that the
$\phi_{\rm ph}=0$ corresponds to the observed phase of photon that
originates from the magnetic pole at $r\rightarrow 0$.

\begin{figure*}
\vspace{0.0in}
  \begin{center}
    \includegraphics[width=0.7\textwidth]{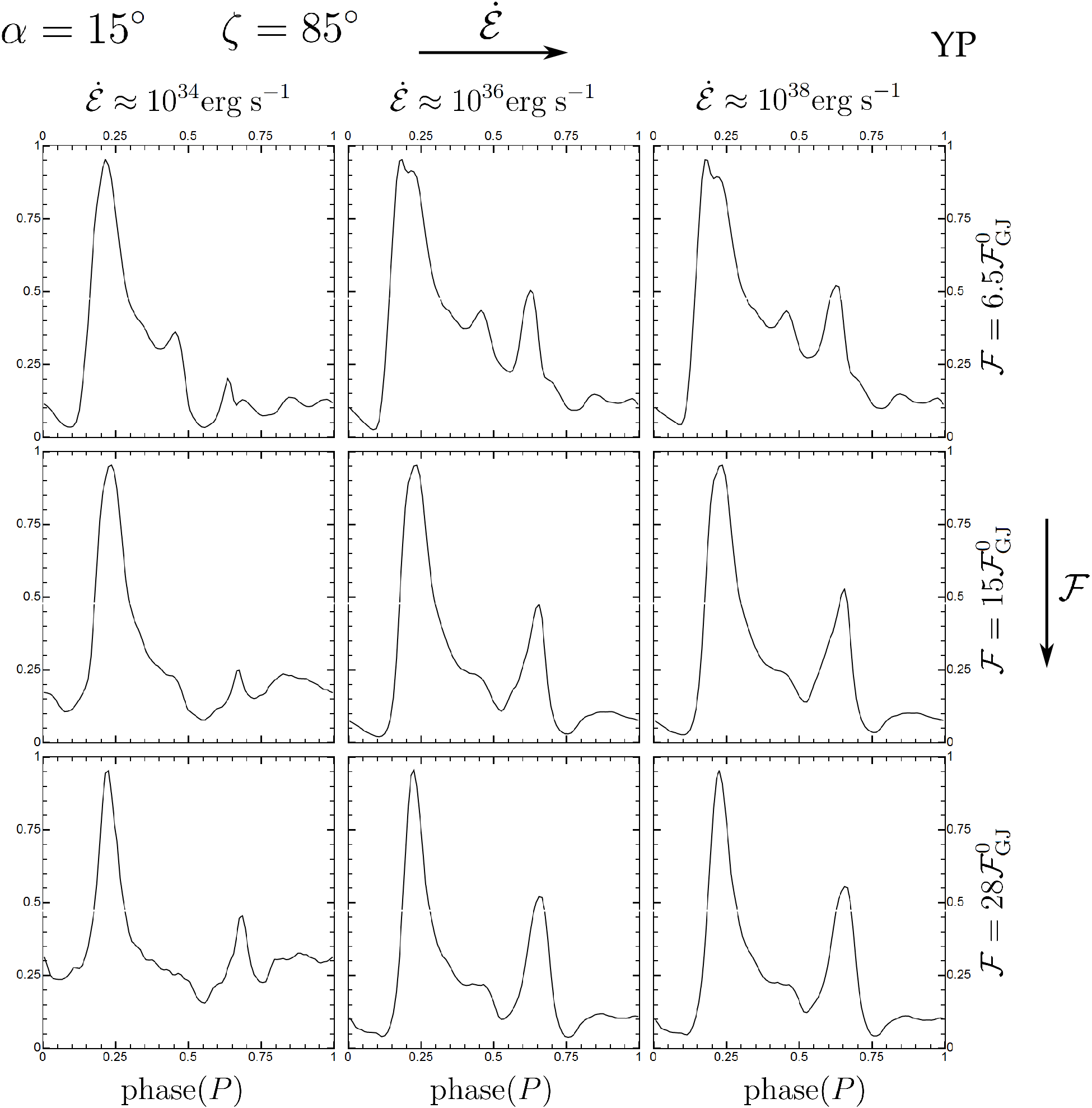}
  \end{center}
  \vspace{0.0in}
  \caption{$\gamma$-ray light-curves for $\alpha=15^{\circ}$.
  The plotted light-curves correspond to the $\zeta$ value ($\zeta=85^{\circ}$)
  indicated by the horizontal white dashed line in Fig.~\ref{fig:012}.}
  \label{fig:0141}
  \vspace{0.0in}
\end{figure*}

The sky-maps have been produced by collecting all the photons
emitted by the particles from $t=1.5P$ to $t=2.0P$ (half a period).
The simulations are in a steady state equilibrium during that time
interval and therefore the photon collection from multiple
time-steps (during this interval) reduces the noise considerably.
The emissivity of each particle is considered $\propto \gamma_{\rm
R}^4 R_{\rm C}^{-2}$ and the corresponding photon emission is along
the particles' velocities. For the determination of the photon
relative time of arrival (i.e. $\phi_{\rm ph}$) we take into account
the corresponding relativistic time delay effects. Thus, $\phi_{\rm
ph}$ reads \citep[see also][]{kalap2014})
\begin{equation}
\label{eq:photonphase} \phi_{\rm ph}=\left(\Omega t_{\rm
S}-\phi_{\pmb{\mathrm{v_{\rm p}}}}-\frac{\pmb{\mathrm{r_{\rm
p}}}\cdot\pmb{\mathrm{v_{\rm p}}}}{\mathrm{v_{\rm
p}}}\frac{1}{R_{\rm
LC}}\right)\negthickspace\negthickspace\negthickspace\mod 2\pi
\end{equation}
where $t_{\rm S}$ is the global simulation time,
$\pmb{\mathrm{v_{\rm p}}}, \pmb{\mathrm{r_{\rm p}}}$ are the
particle velocity and position vectors, and
$\phi_{\pmb{\mathrm{v_{\rm p}}}}$ is the azimuth angle of the
particle velocity $\pmb{\mathrm{v_{\rm p}}}$ with respect to the
magnetic axis at $t_{\rm S}=0$ oriented according to $\pmb{\Omega}$.
The last term in Eq.~\eqref{eq:photonphase} formulates the light
travel time delay.

\begin{figure*}
\vspace{0.0in}
  \begin{center}
    \includegraphics[width=0.7\textwidth]{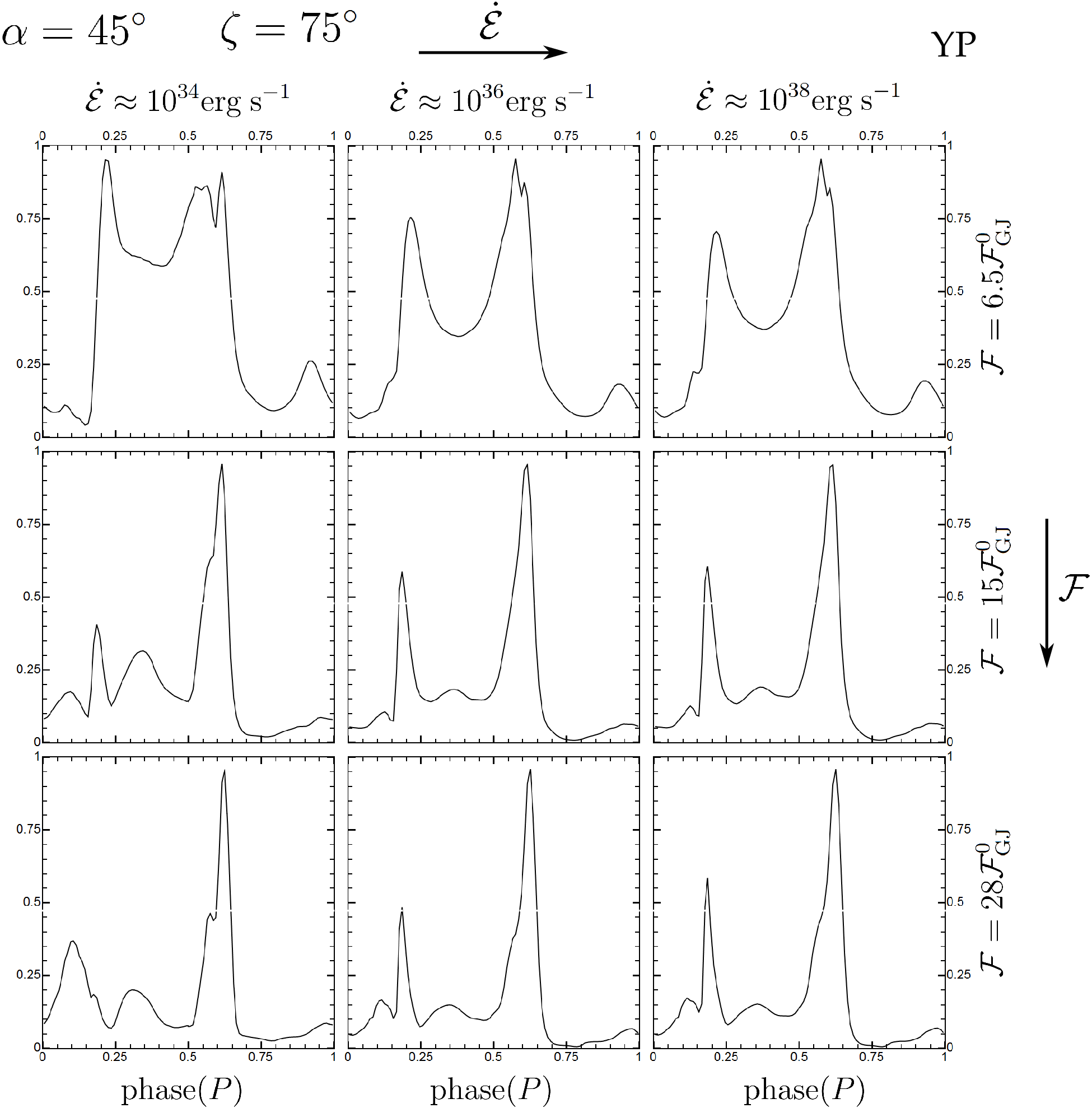}
  \end{center}
  \vspace{0.0in}
  \caption{$\gamma$-ray light-curves for $\alpha=45^{\circ}$.
  The plotted light-curves correspond to the $\zeta$ value ($\zeta=75^{\circ}$)
  indicated by the horizontal white dashed line in Fig.~\ref{fig:013}.}
  \label{fig:0142}
  \vspace{0.0in}
\end{figure*}

The sky-map patterns {(especially towards high $\ir$ values) are
actually} similar to those of the FIDO model \citep[see fig.~17
of~][]{kalap2014}. For high inclination angles the locus of the ECS
deviates considerably from the rotational equator allowing, in
principle, the emission to contribute to very low and very high
$\zeta$ values. However, in our models (of high $\alpha$ values),
similarly to the FIDO models, the main part of the emission does not
come from the high latitude (with respect to the rotational equator)
zones. This makes the corresponding emission more concentrated
around the rotational equator (i.e. around $\zeta=90^{\circ}$).

Actually, in Fig.~\ref{fig:015}, we plot, for the indicated $\alpha$
values and for $\ir=20\ir_{\rm GJ}^0$, the particles that produce
the highest 95\% of the corresponding total $L_{\gamma}$ value. The
individual particle colors represent the corresponding emissivity as
indicated in the color scale. The particle location is always near
the ECS. However, we see that for low $\alpha$ values the particle
distribution is quite uniform while for high $\alpha$ values most of
the emission is produced by particles that are concentrated mainly
closer (with respect to the theoretical extend) to the rotational
equator. This behavior is similar to what had been originally
noticed in the FIDO models \citep[see fig.~19b of~][]{kalap2014}. A
similar behavior is also seen in \cite{2016MNRAS.457.2401C}. {This
implies that the $E_{\rm acc}$, far from the equator, are shorted
out relatively easy (i.e. the corresponding current requirement is
not so strong).} Figure~\ref{fig:015} indicates also that for low
$\alpha$ values the emission is more concentrated near the LC while
for high $\alpha$ values extends to larger distances (see the
distribution of the red color points in Fig.~\ref{fig:015}).

Moreover, Figs.~\ref{fig:012}-\ref{fig:014} indicate that towards
high $\mathcal{F}$ and $\ed$ values the luminous sky-map regions
become narrower. This effect is further enhanced if we take into
account that $\mathcal{F}$ and $\ed$ increase together (see
Fig.~\ref{fig:0105}). This behavior is consistent with the
observations, which show broader $\gamma$-ray light-curves towards
low $\ed$ values of YPs and MPs (2PC).

\begin{figure*}
\vspace{0.0in}
  \begin{center}
    \includegraphics[width=0.7\textwidth]{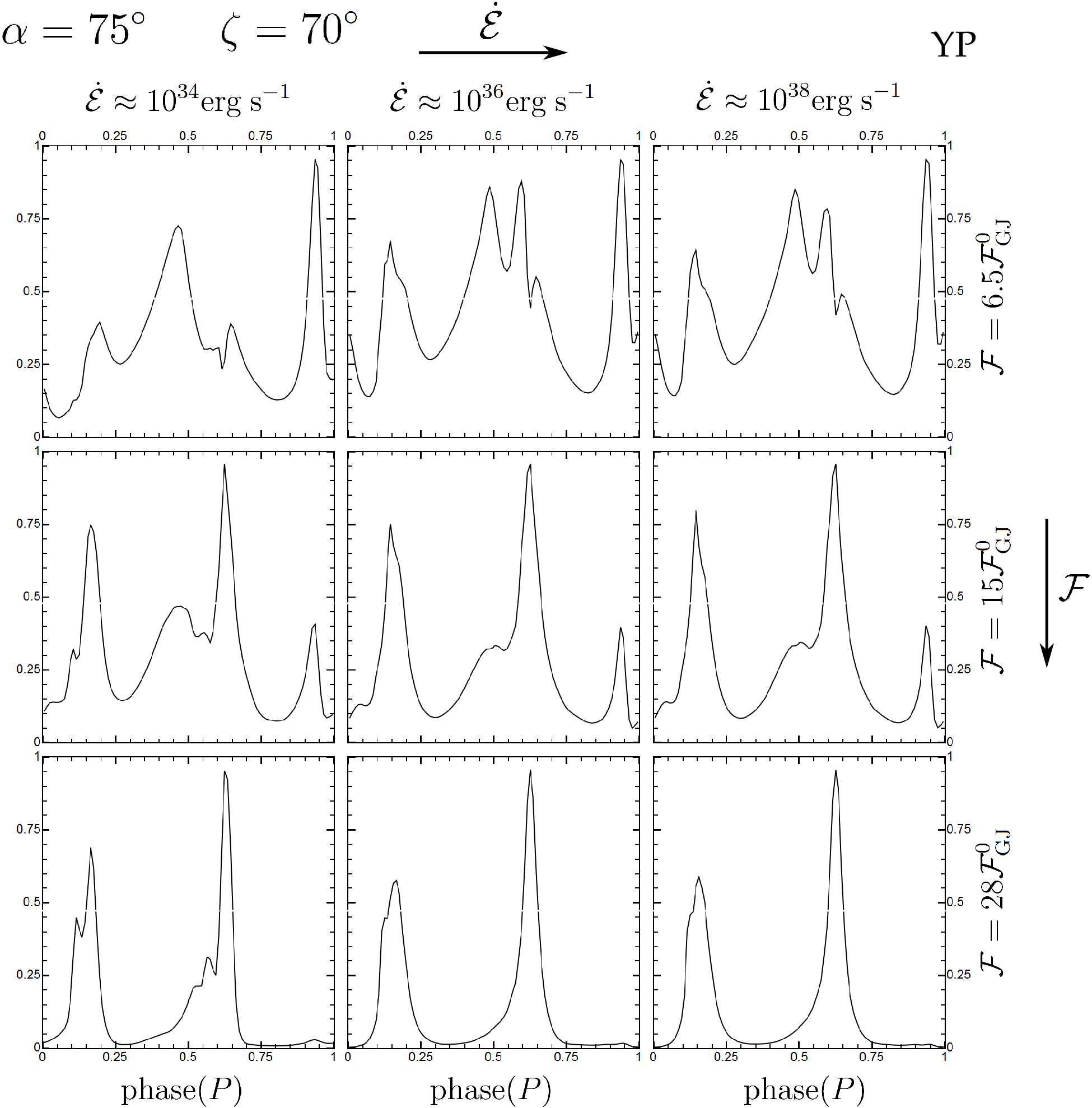}
  \end{center}
  \vspace{0.0in}
  \caption{$\gamma$-ray light-curves for $\alpha=75^{\circ}$.
  The plotted light-curves correspond to the $\zeta$ value ($\zeta=70^{\circ}$)
  indicated by the horizontal white dashed line in Fig.~\ref{fig:014}.}
  \label{fig:0143}
  \vspace{0.0in}
\end{figure*}

In Figs.~\ref{fig:0141}-\ref{fig:0143}, we present, for
demonstration, $\gamma$-ray light-curves for the indicated
$\mathcal{F}$, $\alpha$, $\ed$, and $\zeta$ values. These correspond
to the $\zeta$ values indicated by the dashed white horizontal lines
shown in Figs.~\ref{fig:012}-\ref{fig:014}. The $\gamma$-ray
light-curves corresponding to high $\ir$ values appear very similar
to the observed ones (2PC). As mentioned above, in these cases the
sky-maps and the corresponding $\gamma$-ray light-curves are similar
to the FIDO model ones and so they yield radio-lag values, $\delta$,
consistent with those indicated by \emph{Fermi} (assuming that the
radio emission comes from near the polar cap region).

However, we note that there are cases (mainly for low $\mathcal{F}$
and high $\alpha$ values; see first rows of Fig.~\ref{fig:014} and
\ref{fig:0143}) that the light-curves seem to have more complicated
features (e.g. more than two peaks, considerable inter-peak and
off-peak emission) not seen in the majority of the observed
$\gamma$-ray light-curves. This is mainly due to the adopted
``uniform"\footnote{The term uniform here implies just that the
magnetization threshold in Eq.~\ref{eq:magnthr} is not magnetic
field line dependent.} regulation of the particle injection rate. By
reducing $\mathcal{F}$ uniformly, we start introducing accelerating
electric fields in regions (even inside the closed zone) that
produce additional emission components not always consistent with
the observed features. However, we note that peculiar and unique
features in $\gamma$-ray light-curves exist in some MPs which have
low $\ed$ value and are expected to be pair-starved \citep[i.e. low
$\mathcal{F}$ values;
see~][]{2004ApJ...606.1143M,2014ApJS..213....6J}. Independently of
this, it seems that the construction of realistic low $\mathcal{F}$
models \citep[i.e. weak
pulsars,][]{2013arXiv1303.4094G,2016MNRAS.463L..94C} should be made
more carefully (compared to the aproach followed in this study). Our
results indicate that the observed $\gamma$-ray emission is
regulated mainly by the particle abundance in regions near the ECS.
For the rest of the magnetosphere regions the accelerating electric
components are more easily screened and beyond some point (i.e. some
$\mathcal{F}$ value) no considerable emission is produced. Of
course, the role of these regions might be important for the
lower-energy part of the spectrum (see also the related discussion
in Section~\ref{sec:rescale}).

\begin{figure*}[!tbh]
\vspace{0.0in}
  \begin{center}
    \includegraphics[width=1.0\textwidth]{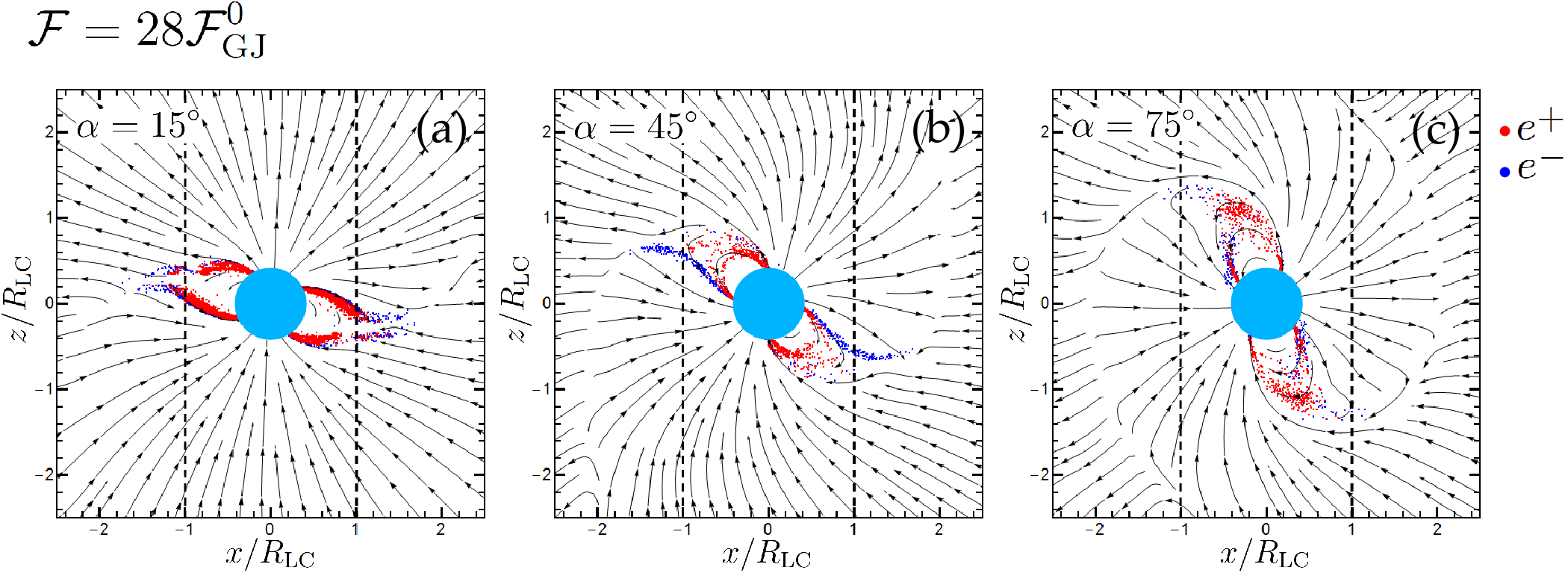}
  \end{center}
  \vspace{0.0in}
  \caption{The origin of the charges (blue $e^-$, red $e^+$) on the
  $\pmb{\mu}-\pmb{\Omega}$ planes for the indicated $\alpha$ values
  that produce the highest 95\% of the total emission for
  $\ir=28\ir_{\rm GJ}^0$.}
  \label{fig:016}
  \vspace{0.0in}
\end{figure*}

In Fig.~\ref{fig:016}, we plot, on the poloidal
$\pmb{\mu}-\pmb{\Omega}$ plane, the origin of the particles (red for
$e^{+}$ and blue for $e^{-}$) that eventually produce the highest
95\% of the observed YP emission, in the case of $\ir=28\ir_{\rm
GJ}^0$. We see that the particles that produce most of the observed
emission are $e^{+}$ and originate along the separatrix. This
particle population goes into the ECS, where it emits. A
lower-energy population that originates mainly by the same locus
regulates self-consistently the corresponding accelerating electric
fields\footnote{As we see in \cite{2017arXiv171003536B} a drifting
particle component can also contribute to the ECS.}. From all the
above, it becomes apparent that an approach similar to what we
followed in \cite{2017ApJ...842...80K}, becomes essential. In
\cite{2017ApJ...842...80K} \citep[see also][]{2014ApJ...781...46C},
we assumed that the dissipative region is only near the ECS beyond
the LC and we applied the finite $\sigma$ values only there. The
equivalent in the PIC simulations would be to apply a rather high
and constant particle injection rate in all the magnetosphere
regions except for the ones indicated in Fig.~\ref{fig:016}. It
seems that the regulation of only this population would provide
results that are in full-agreement with the observations for the
entire range of the parameter space. This approach highlights the
role of the return-current region in the high-energy emission
observed in pulsars, something that, eventually, is expected to be
supported by the (micro) physically motivated pair-creation.

\section{Conclusions \& Discussion} \label{sec:concl}

In this paper, we present 3D kinetic global models of pulsar
magnetospheres. In our study, we use C-3PA, an efficient, 3D
cartesian relativistic Particle-In-Cell code we have developed. In
this first step, we neglect the microphysics of the particle
production and so we apply ad-hoc particle injection rates ($\ir$)
that at least produce pulsar magnetosphere models where the field
structure and the corresponding particle distribution are consistent
with each other.

The particle injection is applied up to large distances (2.5$R_{\rm
LC}$) and is regulated by the local magnetization. From low to high
particle injection rates we cover an entire spectrum of solutions
from near VRD to near FF, respectively. For models of high particle
injection rates ($>1\ir_{\rm GJ}^0$) the particle acceleration takes
place mainly near the region of the equatorial current sheet outside
the LC while the volume of this region decreases with the particle
injection rate.

A main goal of this study is to uncover the properties of the model
$\gamma$-ray emission and how these vary with the particle injection
rate. However, in our PIC simulations the particle Lorentz factors
are (due to numerical limitations) much smaller than those in real
pulsar magnetospheres. Moreover, the relative way the synchrotron
and curvature losses behave, can be quite different for different
particle energy regimes. In our approach, we assume that the
synchrotron losses in the environment of real pulsar magnetospheres
rapidly reduce the corresponding pitch angles and so the synchrotron
emission is not the main component of the observed $\gamma$-ray
component. In our simulations, we use artificially scaled-up
magnetic fields inside the radiation reaction force expressions in
order to reduce the synchrotron cooling times. This affects mostly
the geometry of the trajectories because of the suppressed
gyro-motion. At the same time, along the PIC particle trajectories
we derive realistic particle Lorentz factors by rescaling the fields
and the length and time scales taking always into account the
corresponding energy gain and the curvature radiation reaction
losses. This approach allows the calculation of the corresponding
high-energy emission (i.e. sky-maps, light-curves, spectra).

Using realistic value sets for the stellar surface magnetic field
($B_{\star}$) and period ($P_{\rm s}$) that cover the corresponding
observed value ranges of YPs and MPs, we find the following main
results:
\begin{enumerate}[(a)]
    \item For the same particle injection rate, $\ir$, the cutoff
    energies, $\ec$,
    increase with the spin-down power, $\ed$.

    \item For the same $B_{\star}$ and $P_{\rm s}$ values (i.e. $\ed$) the
    $\ec$ decrease with $\ir$ for high $\ir$ and saturate for low
    $\ir$.

    \item The optimum $\ir$ values that produce $\ec$ within the narrow value range
    observed by \emph{Fermi} increase with $\ed$ and can reach
    up to $10^5$ for both YPs and MPs. These optimum $\ir$ values
    produce also the observed $L_{\gamma}$ behavior
    ($L_{\gamma}\propto \ed$ for low $\ed$ and $L_{\gamma}\propto \sqrt{\ed}$ for high
    $\ed$).

    \item YP and MP models with $\ed\lesssim 10^{33}\rm erg~s^{-1}$ and $\ed\lesssim
    10^{32}\rm erg~s^{-1}$ respectively produce $\ec$ values that are always smaller
    than those observed by \emph{Fermi} and close to the \emph{Fermi} threshold,
    which provides a possible
    explanation for why we do not observe $\gamma$-ray YPs and MPs with
    smaller $\ed$.

    \item The $\gamma$-ray efficiency shows a maximum (at $\ir\approx 1.5\ir_{\rm GJ}^0$) that depends
    on the inclination angle, $\alpha$, and ranges from 13\% to
    35\%. For higher $\ir$ values the corresponding $\gamma$-ray efficiency
    drops considerably.

    \item The higher the $\ed$ and $\ir$, the more narrow the peaks
    in the $\gamma$-ray light-curves are.
\end{enumerate}

For lower $\ir$ values (i.e. $\ir\lesssim 10\ir_{\rm GJ}^0$),
additional features begin appearing (especially for high $\alpha$
values) in the model $\gamma$-ray light-curves that are not always
consistent with the observed ones. This indicates that the decrease
of $\ir$ should not be uniform along all the magnetic field lines
but more focused along the separatricies since these regions are the
main suppliers of the particles that eventually lie in the region
near the ECS. Actually, this is totally consistent with the approach
we have followed for the FIDO models where the decrease of the
plasma conductivity $\sigma$ was made only near the ECS and not in
other places inside or outside the LC \citep{2017ApJ...842...80K}.
The above indicate that the number of particles that eventually
enter the ECS region regulate the observed emission. For this
reason, we have started producing kinetic models using a magnetic
field line dependent particle injection prescription. This approach
allows a much better (i.e. autonomous) control of the particle
population that participates and regulates the area at and around
the ECS. We expect the $\gamma$-ray light-curves of these models to
be consistent with the \emph{Fermi} ones even for the lower particle
injection rates.

In \cite{2017ApJ...842...80K} (FIDO models), we found a dependence
of the plasma conductivity $\sigma$ on $\ed$ while in the present
study (kinetic models) we find a dependence of the global particle
injection rate, $\ir$ on $\ed$ as well. Both these relations
($\sigma$ vs. $\ed$ and $\ir$ vs. $\ed$) are imposed by the
\emph{Fermi} data and more specifically by the requirement that the
models reproduce the observed $\ec$. {Besides the $\ec$-success, the
optimum kinetic models reproduce the trends of the observed total
$\gamma$-ray luminosity as a function of $\ed$. Moreover, the FIDO
models and the kinetic models for high particle injection rates
provide very similar sky-maps and $\gamma$-ray light-curves that are
consistent with the observed $\delta$-$\Delta$ correlation while, as
mentioned above, the $\gamma$-ray light-curves for lower particle
injection rates indicate often that a more careful treatment of the
corresponding particle injection is needed. Nonetheless, the
apparent $\sigma-\ir-\ed$ relation, uncovered by our studies,
connects fundamental macroscopic quantities providing a unique
insight into the understanding of the physical mechanisms behind the
high-energy emission in pulsar magnetospheres.}

Eventually, this successful description needs to be supported by the
microphysical mechanisms that govern the particle creation processes
providing a complete justification for the aforementioned relations.

However, incorporating the pair creation microphysics
\citep[see][]{2013MNRAS.429...20T} in global PIC simulations is not
a trivial task because of the corresponding different length-time
scales and the fact that the PIC energies are much smaller than
those for real pulsars. Nonetheless, trying to expand our studies
toward this direction, we have started exploring {not only} global
magnetosphere models with magnetic field line dependent particle
injection {but also} fully self-consistent models that incorporate a
physically motivated particle injection and we will present our
results in forthcoming papers.

\acknowledgments

We would like to thank an anonymous referee for helpful suggestions
that improved the paper. This work is supported by the National
Science Foundation under Grant No. AST-1616632, by the NASA
Astrophysics Theory Program, by the NASA Astrophysics Data Analysis
Program, and by Fermi Guest Investigator Program. Resources
supporting this work were provided by the NASA High-End Computing
(HEC) Program through the NASA Advanced Supercomputing (NAS)
Facility at NASA Ames Research Center and NASA Center for Climate
Simulation (NCCS) at NASA Goddard Space Flight Center.

\end{document}